\documentclass[preprint,12pt]{elsarticle}

\makeatletter
\newcommand{\dontusepackage}[2][]{
    \@namedef{ver@#2.sty}{9999/12/31}
    \@namedef{opt@#2.sty}{#1}
}

\newcommand{\usepackagesave}[2][{}]{
    \ltx@ifpackageloaded{#2}{}{
        \usepackage[#1]{#2}}
}

\@ifclassloaded{standalone}
{
    \dontusepackage{geometry}
}
{}

\@ifclassloaded{elsarticle}
{
    \dontusepackage{biblatex}
    \dontusepackage{authblk}
}
{}

\@ifclassloaded{beamer}
{
    \dontusepackage{authblk}
    \dontusepackage{titlesec}
    \dontusepackage{footnote}
    \dontusepackage{enumitem}
}
{}

\@ifclassloaded{anstrans}
{
    \dontusepackage{amsthm}
}
{
    
}

\makeatother

\usepackage{standalone}
\usepackage{xifthen}     

\usepackage[english]{babel}	
\usepackage[utf8]{inputenc} 
\usepackage[kerning,spacing,babel]{microtype} 
\usepackage{eepic}      
\usepackage{epic}       
\usepackage[T1]{fontenc} 
\usepackage{lmodern}

\usepackagesave[
    backend=biber,
    natbib=true,
    url=false, 
    doi=true,
    eprint=false
    ]{biblatex}

\usepackage{authblk}		

\usepackage{upgreek}		
\usepackage[makeroom]{cancel}			

\usepackage{isotope}		
\usepackage[version=3]{mhchem}	

\usepackage{xspace}			
\usepackage{icomma}			
\usepackage{chngpage}		
\usepackage[usenames,dvipsnames,svgnames,table]{xcolor}			
\usepackage{enumitem}       
\usepackage{footnote}       

\usepackage[singlelinecheck=false, labelsep=quad]{caption}	
\usepackage{subcaption}		
\usepackage{placeins}   	
\usepackage{float}      	

\usepackage{supertabular}	
\usepackage{booktabs}		
\usepackage{array}			
\usepackage{multirow}		

\usepackage{graphicx}		
\usepackage{wrapfig}		
\usepackage{pgfplots}		
\usepackage{siunitx}

\usepackage{algorithmicx} 
\usepackage{listings} 	
\usepackage{verbatim}   

\usepackage{url}        
\usepackage{import}			
\usepackage{afterpage}  
\usepackage{lscape}     
\usepackage{rotating}   
\usepackage{chngcntr}   
\usepackage{appendix}   

\usepackage{epsfig}     
\usepackage{epsf}

\usepackage{hyperref}

\usepackage{geometry}		
\usepackage{algorithm}  

\usepackage{amsmath}    
\usepackage{amsfonts}   
\usepackage{amssymb}    
\usepackagesave{amsthm}     
\usepackage{bm}  

\usepackage{nameref}		
\usepackage[capitalize,nameinlink]{cleveref}

\makeatletter
\g@addto@macro\@floatboxreset\centering
\newcommand*{\currentname}{\@currentlabelname}
\makeatother

\newlength{\figheight}
\newlength{\figwidth}
\hypersetup{
  linktocpage=true,       
  bookmarks=true,         
  unicode=true,           
  pdftoolbar=true,        
  pdfmenubar=true,        
  pdffitwindow=false,     
  pdfstartview={FitH},    
  pdfnewwindow=true,      
  colorlinks=false,       
  linkcolor=red,          
  citecolor=green,        
  filecolor=magenta,      
  urlcolor=cyan,          
  pdfborder={0 0 0},      
}

\crefname{appsec}{Appendix}{Appendices}

\pgfplotsset{compat=1.14}

\pgfplotsset{
    log x ticks with fixed point/.style={
        xticklabel={
            \pgfkeys{/pgf/fpu=true}
            \pgfmathparse{exp(\tick)}
            \pgfmathprintnumber[fixed relative, precision=3]{\pgfmathresult}
            \pgfkeys{/pgf/fpu=false}
        }
    },
    log y ticks with fixed point/.style={
        yticklabel={
            \pgfkeys{/pgf/fpu=true}
            \pgfmathparse{exp(\tick)}
            \pgfmathprintnumber[fixed relative, precision=3]{\pgfmathresult}
            \pgfkeys{/pgf/fpu=false}
        }
    }
}

\newcommand{\apriori}{\textit{a priori}\xspace}

\newcommand{\bracet}[1]{{\left\{#1 \right\}}}

\newcommand{\del}{\vec{\nabla}}

\newcommand{\vd}{\mathbf{\cdot}}
\renewcommand{\div}{\del \vd }

\newcommand{\norm}[2][{}]{\lVert#2\rVert_{#1}}
\newcommand{\e}[1]{\mathrm{e}^{#1}}

\newcommand{\avg}[1]{\ensuremath{\overline{#1}}\xspace}
\newcommand{\linear}[1]{\ensuremath{\widetilde{#1}}\xspace}
\newcommand{\linearv}[1]{\ensuremath{\vec{\widetilde{#1}}}\xspace}

\newcommand{\jacobian}{\mat{\mathbb{J}}}

\newcommand{\half}[1][1]{\frac{#1}{2}}

\newcommand{\mat}[1]{\ensuremath{\mathbf{#1}}}

\newcommand{\dd}[2][]{\frac{\partial #1}{\partial #2}}

\newcommand{\ddt}[1][]{\dd[#1]{t}}

\newcommand{\dx}[1][x]{\,d#1}

\newcommand{\dt}{\dx[t]}

\newcommand{\domg}{\dx[\Omega]}

\newcommand{\intsp}{\int_{4\pi}}

\newcommand{\normal}{\ensuremath{\hat{n}}\xspace}

\newcommand{\plank}{\ensuremath{h}\xspace}
\newcommand{\boltzmann}{\ensuremath{k_\mathrm{B}}\xspace}

\newcommand{\cv}{\ensuremath{c_{v}}\xspace}
\newcommand{\density}{\ensuremath{\rho}\xspace}

\newcommand{\frequency}{\ensuremath{\nu}\xspace}
\newcommand{\pos}{\ensuremath{\vec{x}}\xspace}

\newcommand{\sn}[1][N]{\ensuremath{S_#1}\xspace}
\newcommand{\pn}[1][N]{\ensuremath{P_#1}\xspace}

\newcommand{\addindex}[1]{\ifthenelse{\isempty{#1}}{}{,{#1}}}

\newcommand{\addgroup}[1]{\ifthenelse{\isempty{#1}}{}{_{#1}}}
\newcommand{\direction}{\ensuremath{\hat{\Omega}}\xspace}

\newcommand{\drift}[1][]{\ensuremath{\hat{D}\addgroup{#1}}\xspace}

\newcommand{\xslabel}[2][]{\ifthenelse{\isempty{#1}}{\mathrm{#2}}{\mathrm{#2},#1}}

\newcommand{\opacity}[1][]{\ensuremath{\sigma_{\addgroup{#1}}}\xspace}
\newcommand{\opacityP}[1][]{\ensuremath{\sigma_{\mathrm{P}\addgroup{#1}}}\xspace}
\newcommand{\opacityE}[1][]{\ensuremath{\sigma_{\mathrm{E}\addgroup{#1}}}\xspace}
\newcommand{\opacityR}[1][]{\ensuremath{\sigma_{\mathrm{R}\addgroup{#1}}}\xspace}

\newcommand{\weight}[1][]{\ensuremath{\mathrm{w}\addgroup{#1}}\xspace}

\hypersetup{
    pdftitle={A multi-dimensional, moment-accelerated deterministic particle method for time-dependent, multi-frequency thermal radiative transfer problems},
    pdfauthor={Hans R. Hammer},
    pdfsubject={Deterministic particle method for thermal radiation transport},
    pdfkeywords={thermal radiation transport,}
}

\newcommand{\HO}{\mathrm{\,{ HO}}}
\newcommand{\LO}{\mathrm{\,{  LO}}}

\renewcommand{\drift}{\ensuremath{\vec{\gamma}^\HO}\xspace}
\newcommand{\resSource}{\ensuremath{\mathcal{S}^\HO}\xspace}
\newcommand{\ccenter}{\ensuremath{\pos_{\mathrm{C},i}}\xspace}
\renewcommand{\vec}[1]{\bm{#1}} 

\renewcommand{\normal}{\ensuremath{\hat{\vec{n}}}\xspace}
\renewcommand{\direction}{\ensuremath{\hat{\vec{\Omega}}}\xspace}

\crefname{appsec}{Appendix}{Appendices}

\usetikzlibrary{patterns}
\usepgfplotslibrary{external} 
\begin{document}
    \begin{frontmatter}
        
        \title{A multi-dimensional, moment-accelerated deterministic particle method for time-dependent, multi-frequency thermal radiative transfer problems}
        
        \author[t3]{Hans Hammer\corref{cor1}}
        \ead{hrhammer@lanl.gov}
        \author[t3]{HyeongKae Park}
        \ead{hkpark@lanl.gov}
        \author[t5]{Luis Chac\'{o}n}
        \ead{chacon@lanl.gov}
        
        \cortext[cor1]{Corresponding author}
        \address[t3]{Fluid Dynamics and Solid Mechanics (T-3)\\ Los Alamos National Laboratory\\ Los Alamos, NM, 87545}
        \address[t5]{Applied Mathematics and Plasma Physics (T-5)\\ Los Alamos National Laboratory\\ Los Alamos, NM, 87545}
        
        \begin{abstract}
               Thermal Radiative Transfer (TRT) is the dominant energy transfer mechanism in high-energy density physics with applications in inertial confinement fusion and astrophysics. The stiff interactions between the material and radiation fields make TRT problems challenging to model. In this study, we propose a multi-dimensional extension of the deterministic particle (DP) method. The DP  method combines aspects from both particle and deterministic methods. If the emission source is known \apriori, and no physical scattering is present, the intensity of a particle can be integrated analytically. This introduces no statistical noise compared to Monte-Carlo methods, while maintaining the flexibility of particle methods. The method is closely related to the popular method of long characteristics. The combination of the DP-method with a discretely-consistent, nonlinear, gray low-order system enables an efficient solution algorithm for multi-frequency TRT problems. We demonstrate with numerical examples that the use of a linear-source approximation based on spatial moments improves the behavior of our method in the thick diffusion limit significantly.
        \end{abstract}
        
        
    \end{frontmatter}
    
\section{Introduction} \label{sec:introduction}

    Many applications in astrophysics and plasma physics require the simulation of high-energy density phenomena. In these regimes, thermal X-ray radiation is the dominant mechanism of energy transfer. Thermal radiative transfer (TRT) is described by a system of stiff, nonlinear equations, and therefore difficult to model.
    
    Two distinct methods are typically used, namely, the deterministic transport method and stochastic particle methods. Both methods have their advantages and disadvantages. A popular particle method is Implicit Monte Carlo (IMC)\cite{fleck_implicit_1971}, which is based on a linearization of the reemission physics.  The particle approach gives IMC flexibility for complicated geometries. The most common deterministic methods are the discrete ordinance or \sn method and the method of characteristics (MOC). Both are frequently used for neutron transport calculations in reactor physics. The advantage of \sn and MOC over IMC is that they do not show statistical noise, and they are able to obtain the asymptotic diffusion limit~\cite{larsen_asymptotic_1987,larsen_asymptotic_1989,larsen_asymptotic_1983} in a relatively straightforward manner~\cite{adams_discontinuous_2001,adams_characteristic_1998}.
    
     A newly proposed deterministic particle (DP)  method~\cite{park_multigroup_2019} combines these two approaches. If the emission source is known \apriori, the intensity of a particle can be integrated analytically. In contrast to  Monte-Carlo (MC) methods, our method does not feature any randomness (with the possible exception of particle initialization), and hence does not show stochastic noise, while maintaining the flexibility of particle methods. This method is similar to MOC in space and time proposed by \citet{pandya_method_2009}. However, our method is particle based, and therefore does not require fixed tracks.
     
     When this DP solver is cast in a high-order, low-order (HOLO) algorithmic framework~\cite{chacon_multiscale_2017}, a discretely-consistent, gray, nonlinear low-order (LO) system provides a well-informed emission source, acts as an interface with other physics in a multi-physics setting, and provides algorithmic acceleration~\cite{chacon_multiscale_2017}. Because of the presence of the consistent, nonlinear LO system, the emission source is known and the contribution from the absorption-emission physics can be analytically integrated along a particle's trajectory. This implicit treatment of the absorption-emission physics removes the explicit time step constraint, allowing us to choose the time step size based on accuracy, not stability.
     
      The DP-HOLO method has been recently demonstrated in one dimension~\cite{park_multigroup_2019}. In this work we present the extension of the DP-HOLO method to multiple dimensions using a ray-tracing approach and a linear-source reconstruction scheme that preserves the diffusion limit. The flat-source approximation requires many cells in optically thick materials to model the changes in the temperature and the emission source sufficiently well. Introducing a linear reconstruction reduces the number of cells necessary to obtain a good representation of the source. We use the method proposed by \citet{ferrer_linear_2016,ferrer_linear_2018} for MOC to obtain the linear representation of tallies. \citet{adams_characteristic_1998} showed that linear source representation on a triangular mesh, and bi-linear on orthogonal, rectangular meshes are required to obtain the thick diffusion limit. For general polygons, piecewise linear basis functions are required as shown by \citet{pandya_long_2011}, which again reduce to linear functions on triangles. Therefore, a linear-source reconstruction is also necessary to obtain the asymptotic diffusion limit \cite{larsen_asymptotic_1987,larsen_asymptotic_1989} with our method.
      
      The reminder of the paper is structured as follows. We first introduce our method in \cref{sec:method}, with the LO solver first followed by the HO solver. We then present numerical results to show the capabilities of this method in \cref{sec:results}. We chose the Tophat~\cite{gentile_implicit_2001} problem to demonstrate the effects of the linear source approximation and a planar Hohlraum problem~\cite{mcclarren_robust_2010,brunner_forms_2002} to demonstrate the effects of different quadratures and random particle initialization. We compare our results to Capsaicin (which employs an $S_N$ implementation)~\cite{thompson_capsaicin:_2006}. Finally, we show results for runtime and convergence studies of the algorithm before we conclude.
    
\section{Method} \label{sec:method}

    Thermal radiative transfer without physical scattering can be described by the following system of equations
    \begin{align}
       \frac{1}{c} \ddt[I] + \direction \vd \del I + \opacity I &= \opacity B \\
       \density \cv \ddt[T] &= \int_{0}^{\infty}\intsp  \left(\opacity I - \opacity B\right) \domg \dx[\frequency]
    \end{align}
    where \(I\left(\pos, \direction, \frequency, t\right)\) is the specific radiation intensity at position \pos traveling in along direction \direction with the speed of light \(c\) and frequency \frequency at time \(t\). Here, 
    is the frequency integrated (gray) radiation energy density, \(\opacity\left(\pos, \frequency, T\right)\) is the opacity at temperature \(T\), \density is the host material's density, and \cv is its specific heat capacity. The emission spectrum is defined by the Planck function
    \begin{equation}
        B\left(\frequency, T\right) \equiv \frac{2\plank \frequency^3}{c^2}\frac{1}{\e{{\plank\frequency}/{\boltzmann T}} - 1},
    \end{equation}
    where \plank denotes the Planck constant and \boltzmann the Boltzmann constant.
    
    Solving the TRT system is difficult due to the high dimensionality of the phase-space and the stiff, nonlinear coupling between the radiation field and the material temperature described by the absorption and emission physics. 
    
    The  DP algorithm uses a high-order (HO), multi-frequency transport solve combined with a gray low-order solver (LO). For simplicity, we will describe the basic algorithm for the gray case and will extend it to the multi-frequency case afterwards. The gray TRT equation is
    \begin{equation} \label{eq:method:transport}
        \frac{1}{c} \ddt[I] + \direction \vd \del I + \opacity I = Q^\LO\left(\pos, t\right).
    \end{equation}
    The DP method requires an \apriori known emission source
        \begin{equation} \label{eq:method:emission_source}
            Q^\LO\left(\pos,t\right) = \frac{\opacity acT^4\left(\pos, t\right)}{4\pi}
        \end{equation} 
    for the integration along the particle trajectory, where \(a\) is the radiation constant. Here, the LO superscript indicates that the source is evaluated from low-order quantities, obtained from a recently developed iterative, moment-based HOLO algorithm~\cite{park_consistent_2012,park_efficient_2013,chacon_multiscale_2017}. The LO system is defined by taking the first two angular moments of \cref{eq:method:transport} together with the material temperature equation,
    \begin{subequations} \label{eq:method:lo_system_consistent}
        \begin{align}
        \label{eq:method:lo_system_consistent:balance}
        \ddt[E^\LO] + \div \vec{F}^\LO + \opacity c E^\LO &= \opacity ac T^{4} + \resSource                           \\
        \label{eq:method:lo_system_consistent:flux}
        \frac{1}{c} \ddt[\vec{F}^\LO] +  \frac{c}{3} \del E^\LO + \opacity \vec{F}^\LO &= \drift cE^\LO \\
        \label{eq:method:lo_system_consistent:temperature}
        \density \cv \ddt[T] + \opacity ac T^4 &=  \opacity cE^\LO,
        \end{align}
    \end{subequations}  
    where the HO superscript indicates that the quantity is evaluated using HO quantities,
            \begin{equation} \label{eq:method:radiation_energy}
                E\left(\pos, t\right) \equiv \frac{1}{c} \int_{0}^{\infty} \intsp I \domg \dx[\frequency]
            \end{equation}
        is the frequency integrated (gray) radiation energy density, and
    \begin{equation} \label{eq:method:flux}
        \vec{F}\left(\pos, t\right) \equiv \int_{0}^{\infty} \intsp \direction I \domg \dx[\frequency]
    \end{equation} 
    is the gray radiative flux.  We used the standard \(\pn[1]\) closure in \cref{eq:method:lo_system_consistent:flux} instead of the more consistent Eddington tensor closure~\cite{goldin_quasi-diffusion_1964}. This introduces inconsistencies between the HO and LO descriptions, in addition to inconsistencies in the discretization. Adding the consistency term \(\drift\) to \cref{eq:method:lo_system_consistent:flux} will correct for transport effects and the mismatch in truncation errors. While \cref{eq:method:lo_system_consistent:balance} is exact in the continuum, we advance that we will need to correct for mismatches between HO and LO temporal discretizations, and add the residual source term 
    \begin{equation}
        \resSource = \ddt[E^\HO] \Big|_\mathrm{LO} - \ddt[E^\HO] \Big|_\mathrm{HO}.
    \end{equation} Here, the subscript LO indicates that we apply the LO discrete temporal derivative scheme, and similarly with the HO subscript. Both the LO and the HO methods conserve energy.
    
    Both \drift and \resSource are evaluated from the HO solution, and their specific form depends on the discretization used. Details will be discussed in the next section. 
    
\subsection{Low-order solver} \label{sec:method:lo}
    The integral balance equation for \cref{eq:method:lo_system_consistent:balance} obtained by integrating in time and cell volume is
    \begin{equation} \label{eq:method:lo:balance}
        \frac{\avg{E}^\LO_{i,n+\half} - \avg{E}^\LO_{i,n-\half}}{\Delta t_n} + \sum_{j \in i} \normal_{ij} \vd \normal_{j}\frac{\avg{F}^\LO_{j,n} A_j}{V_i} + \opacity[i,n] c \avg{E}^\LO_{i,n} - \opacity[i,n] ac\avg{T_{i,n}^4} = 0
    \end{equation}
    where \(\Delta t_n = t_{n+\half} - t_{n-\half}\) is the time step size for step \(n\), \(V_i\) the cell volume, \(A_j\) the surface area, \(\normal_j\) is the global unit normal of surface \(j\) and \(\normal_{ij}\) is the unit normal of surface \(j\) pointing outwards from cell \(i\). The sign of the product \(\normal_j \vd \normal_{ij}\) indicates if the flux is outgoing or incoming. The bar notation denotes a spatial average over a cell or surface. Note that we use half indices for end-of-time-step variables and integer indices for time averaged quantities. Therefore, we find
    \begin{subequations} \label{eq:method:lo:ho_quantities}
        \begin{align}
        \label{eq:method:lo:ho_quantities:E_end}
            \avg{E}_{i,n+\half} &\equiv \frac{1}{cV_i} \int_{V_{i}} \intsp I(\pos, \direction, t_{n+\half}) \domg \dx[V] \\
            \label{eq:method:lo:ho_quantities:E_avg}
            \avg{E}_{i,n} &\equiv \frac{1}{c\Delta t_{n}V_i} \int_{t_{n-\half}}^{t_{n+\half}} \int_{V_{i}} \intsp I\left(\pos, \direction, t\right) \domg \dx[V] \dt \\
            \label{eq:method:lo:ho_quantities:F}
            \avg{F}_{j,n} &\equiv \frac{1}{\Delta t_{n}A_j} \int_{t_{n-\half}}^{t_{n+\half}} \int_{A_j} \intsp \normal_j \vd \direction I\left(\pos, \direction, t\right) \domg \dx[A] \dt \\
            \avg{T^4}_{i,n} &\equiv \frac{1}{\Delta t_{n}V_i} \int_{t_{n-\half}}^{t_{n+\half}} \int_{V_{i}} T^4\left(\pos, t\right) \dx[V] \dt
        \end{align}
    \end{subequations}
    The time-discrete LO equation cannot update simultaneously quantities defined at \(n\) and \(n+\half\), and therefore we replace end-of-time-step quantities with time step average quantities by adding a residual source term:
    \begin{equation} \label{eq:method:lo:balance_discrete}
        \frac{\avg{E}^\LO_{i,n} - \avg{E}^\LO_{i,n-1}}{\Delta t_n} + \sum_{j \in i} \normal_{ij} \vd \normal_{j} \frac{\avg{F}^\LO_{j,n} A_j}{V_i} + \opacity[i,n] c \avg{E}^\LO_{i,n} - \opacity[i,n] ac\avg{T^4_{i,n}} = \resSource_{i,n},
    \end{equation}
    where the residual source is defined as:
    \begin{equation} \label{eq:method:lo:residual_source}
        \resSource_{i,n} \equiv \frac{\avg{E}^\HO_{i,n} - \avg{E}^\HO_{i,n-1}}{\Delta t_n} - \frac{\avg{E}^\HO_{i,n+\half} - \avg{E}^\HO_{i,n-\half}}{\Delta t_n}.
    \end{equation}
    The discrete equation for the flux \(\avg{F}_{j,n}\) across surface \(j\) using the method proposed by \citet{park_multigroup_2019} can be written as
    \begin{equation}
        \frac{1}{c} \frac{\avg{F}^\LO_{j,n} - \avg{F}^\LO_{j,n-1}}{\Delta t_{n}} + \frac{c}{3}\frac{\avg{E}^\LO_{j+\half,n} - \avg{E}^\LO_{j-\half,n}}{\Delta x_{j}} + \opacity[j,n] \avg{F}^\LO_{j,n} = \gamma^{+\HO}_{j,n} c\avg{E}^\LO_{j+\half,n} - \gamma^{-\HO}_{j,n} c\avg{E}^\LO_{j-\half,n}.
    \end{equation}
    where the indices \(j\pm\half\) denote the cells adjacent to surface \(j\), \(\Delta x_j\) is the characteristic length between these cells, and \opacity[j,n] is a weighted opacity at the surface. The consistency terms are given by \cite{park_consistent_2012}
    \begin{subequations} \label{eq:method:lo:consistency}
        \begin{align}
        \label{eq:method:lo:consistency:+}
        \gamma^{+\HO}_{j,n}  &= \frac{1}{c \avg{E}^\HO_{j+\half,n}}
        \left(\frac{1}{c} \frac{f^{+\HO}_{j,n} - f^{+\HO}_{j,n-1}}{\Delta t_n} + \frac{c}{6}\frac{\avg{E}^\HO_{j+\half,n} - \avg{E}^\HO_{j-\half,n}}{\Delta x_j} + \opacity[j,n] f^{+\HO}_{j,n}\right) \\
        \label{eq:method:lo:consistency:-}
        \gamma^{-\HO}_{j,n} &= \frac{1}{ c \avg{E}^\HO_{j-\half,n}}
        \left(\frac{1}{c} \frac{f^{-\HO}_{j,n} - f^{-\HO}_{j,n-1}}{\Delta t_n} - \frac{c}{6}\frac{\avg{E}^\HO_{j+\half,n} - \avg{E}^\HO_{j-\half,n}}{\Delta x_j} + \opacity[j,n] f^{-\HO}_{j,n} \right)
        \end{align}
    \end{subequations}
    where the partial fluxes are defined from the HO solution with respect to the global surface normal \(\normal_j\) as
    \begin{subequations} \label{eq:method:lo:partial_flux}
        \begin{align}
        \label{eq:method:lo:partial_flux:+}
        f^+_{j,n} &\equiv \frac{1}{\Delta t_n} \int_{t_{n-\half}}^{t_{n+\half}} \int_{A_j} \int_{\direction \vd \normal_j > 0} \normal_j\vd \direction I\left(\pos, \direction, t\right) \domg \dx[A] \dt \\
        \label{eq:method:lo:partial_flux:-}
        f^-_{j,n} &\equiv - \frac{1}{\Delta t_n} \int_{t_{n-\half}}^{t_{n+\half}} \int_{A_j} \int_{\direction \vd \normal_j < 0} \normal_j\vd \direction I\left(\pos, \direction, t\right) \domg \dx[A] \dt.
        \end{align}
    \end{subequations}

    Finally, the flux across the boundary surface \(j\) is 
    \begin{equation}
        \avg{F}_{j,n} = f^+_{j,n} - f^-_{j,n} 
        = \left(1 - \alpha_{j}\right) f^+_{j,n} - f'^-_{j,n} 
    \end{equation}
    where we assume the normal points outwards, and \(\alpha_{j} \in \left[0,1\right]\) is the reflection or albedo factor. It allows the user to define boundaries as vacuum, partially or fully reflecting. The incoming boundary flux without reflection is
    \begin{equation}
        f'^-_{j,n} = f^-_{j,n}  - \alpha_{j} f^+_{j,n} = \frac{acT_{\mathrm{BC},j}^4}{4}.
    \end{equation}
    Defining the boundary factors
    \begin{subequations}
        \begin{align}
            \kappa^{+\HO}_{j,n} &\equiv \frac{f^{+\HO}_{j,n}}{c\avg{E}^\HO_{j-\half,n}} \\
            \kappa^{-\HO}_{j,n} &\equiv 
                \begin{cases}
                    0 & T_{\mathrm{BC},j} = 0 \\
                    \frac{f^{-\HO}_{j,n} - \alpha_{j}f^{+\HO}_{j,n}}{acT_{\mathrm{BC},j}^4} & T_{\mathrm{BC},j} > 0
                \end{cases}
        \end{align}
    \end{subequations}
    gives the LO boundary condition
    \begin{equation}
        \avg{F}^\LO_{j,n} = \left(1 - \alpha_{j}\right) \kappa^{+\HO}_{j,n} c\avg{E}^\LO_{i-\half,n} -\kappa^{-\HO}_{j,n} acT_{\mathrm{BC}, j}^4.
    \end{equation}
    In these equations, \resSource, \(\gamma^{\pm\HO}_{j,n}\) and \(\kappa^{\pm\HO}_{j,n}\) are evaluated from the HO system using \cref{eq:method:lo:ho_quantities,eq:method:lo:partial_flux}. This gives discrete consistency between the LO and HO system. The LO-system is solved using a Newton-Krylov method with non-linear elimination~\cite{park_consistent_2012}. The details are given in \cref{sec:appendix:lo_solution}. \par
    
    Finally, the temperature equation, \cref{eq:method:lo_system_consistent:temperature}, is discretized as
    \begin{equation}
        \density\cv \frac{\avg{T}^\LO_{i,n} - \avg{T}^\LO_{i,n-1}}{\Delta t_n} + \opacity[i,n] ac \avg{T^4_{i,n}} - \opacity[i,n] c \avg{E}^\LO_{i,n} = 0.
    \end{equation}
    
\subsection{High-order solver} \label{sec:method:ho}
    The high-order solver is a particle-based ray-tracing algorithm. In particle-based methods (e.g. Monte Carlo) the angular intensity \(I\) is represented as a collection of \(P\) particles with their specific intensity \(I^\HO_p\) as
    \begin{equation} \label{eq:method:ho:particle_field}
        I^\HO\left(\pos, \direction, t\right) = \sum_{p = 1}^{P} \weight[p] I^\HO_p\left(t\right) \delta\left(\pos - \pos_p\left(t\right)\right)\delta\left(\norm{\direction - \direction_{p}\left(t\right)}\right)
    \end{equation}
    where \(\pos_p\left(t\right)\), \(\direction_{p}\left(t\right)\) are the spatial position and direction of particle \(p\) at time \(t\). The particle phase-space volume \weight[p] is the analog to the track width of MOC methods, an integral weight factor. Details of its calculation are given later.  The evolution equation of the particle intensity can by found by multiplying \cref{eq:method:transport} by \(\delta\left(\pos - \pos_p\left(t\right)\right)\delta\left(\norm{\direction - \direction_{p}\left(t\right)}\right)\) and integrating over the phase-space to obtain
    \begin{equation} \label{eq:method:ho:evolution}
        \frac{1}{c}\frac{\dx[I^\HO_p]}{\dt} + \opacity I^\HO_p = Q^\LO\left(x_p, t\right)
    \end{equation}
    where \(Q^\HO\left(x, t\right)\) is the \apriori known emission source \cref{eq:method:emission_source}. The formal solution for \(I^\HO_p\) along the characteristic for particle \(p\) in cell \(i\) at time step \(n\) is
    \begin{equation} \label{eq:method:ho:characteristic}
        I^\HO_{p,i,n}\left(t\right) = I^\HO_{p,i,n}\left(t_0\right) \e{-\int_{t_0}^{t} \opacity c\dt'}
        + \int_{t_0}^{t} \e{-\int_{t'}^{t}\opacity[i,n] c\dt''} Q^\LO_{i,n}\left(\pos_{p}\left(t'\right), t'\right) c\dt'.
    \end{equation} This equation, and its first spatial moment, can be integrated analytically when one considers a linear emission source (as will be the case here), and that particle trajectories are straight. We consider the particle initialization and trajectory computation next.

\subsubsection{Particle initialization} \label{sec:appendix:particle_initalization}

	The particles are initialized on a per-cell basis. In each cell \(i\), a number of points are selected as starting points for the particles. The points are found by increasingly refining the cell up to a specified level. Each triangle or rectangle cell is divided into four subcells, until the level of requested refinement is reached. After the cell is refined, the particles are initialized at the center point of each subcell \(\zeta\) with the volume \(V_\zeta\). The particles at each point are launched in the direction of a given angular quadrature with corresponding weights \(\bracet{\direction_m, \omega_m}_{m = 1}^{M}\). This quadrature can be deterministic or random. The particle phase-space factor is 
	\begin{equation}
	\weight[p] \equiv V_\zeta \omega_m
	\end{equation}
	and must satisfy
	\begin{equation}
	\sum_{p \in i} \weight[p] = 4\pi V_i.
	\end{equation}

	In this work, we assume initial conditions that are isotropic in angle and Planckian in frequency. Therefore the initial radiation energy density can be described with
	\begin{equation}
	\avg{E}_{i,\half} = aT_{i,\half}^4
	\end{equation}
	where \(T_{i,\half}\) is the initial temperature at \(t = 0\). Using \cref{eq:method:ho:particle_field} and the definition of the radiation energy density, \cref{eq:method:radiation_energy}, we find
	\begin{equation}
	I^\HO_{p,i,\half} = \frac{acT_{i,\half}^4}{4\pi}.
	\end{equation}

\subsubsection{Particle trajectory}
    Since no accelerating forces affect the particles, their trajectory can be simply expressed as a straight ray
    \begin{equation} \label{eq:method:ho:ray}
        \pos_{p}\left(t\right) = \pos_{0,p} + ct\direction_{p}
    \end{equation}
    from \(\pos_{0,p} = \pos_{p}\left(t_0\right)\) in direction \(\direction_{p}\). In curvilinear geometry, local orthogonal coordinates can be used~\cite{park_multigroup_2018-1}. The distance a particle travels within a timestep is \(s_p = c \Delta t\). The particle movement is subdivided  by intersections with cell surfaces. Each subdivision produces a straight track, contained within one cell with constant material properties. After each track, either the cell changes, the particle is reflected at a boundary or the end of the time step is reached. The calculation of intersections is a common problem in computational geometry or graphics applications and many efficient algorithms for all types of surfaces can be found in literature~\cite{glassner_introduction_1989}. We limit our mesh to cells that are strictly convex with planar surfaces. In this case, the intersection between a ray originating from within the cell and the surface of the cell is the shortest positive distance to all of the surfaces. This approach allows us to avoid costly vertex comparisons to determine which surface the particle goes through.
    
    An infinite, planar surface in three dimensions is fully described by the implicit definition
    \begin{equation} \label{eq:method:ho:surface}
        \normal \cdot \pos - b = 0
    \end{equation}
    where \( \normal \in \mathbb{R}^3\) is the normal of the surface and \(b \in \mathbb{R}\) is the offset. With \cref{eq:method:ho:ray} the distance the particle has to travel to cross surface \(j\) from its origin \(\pos_{0,p,i,n}\) in cell \(i\) at time step \(n\) can be found as
    \begin{equation} \label{eq:method:ho:intersection}
        s_{j,p,i,n} = \frac{b_{j} - \normal_{j} \cdot \pos_{0,p,i,n}}{\normal_{j} \cdot \direction_p}.
    \end{equation}
    The surfaces are extended to infinity beyond the limits of the cell. Therefore, there is an intersection with the ray, if
    \begin{equation}    
        \normal_j \cdot \direction_p \ne 0,
    \end{equation}
    otherwise the surface and the ray are parallel. The relevant intersection of the ray is then the intersection with the smallest positive distance
    \begin{equation}
        s_{p,i,n} = \min\limits_{j} s_{j,p,i,n} \qquad \text{for}~ s_{j,p,i,n} > 0.
    \end{equation}
    Special care is necessary if the particle hits a corner as detailed in \cref{sec:appendix:particle_corner}. The time it takes the particle to reach the surface is
    \begin{equation}
        \Delta t_{s,i} = \frac{s_{p,i,n}}{c}
    \end{equation}
    and it must be smaller than the remaining time \(\Delta t_p\) in the time step. Otherwise the particle cannot reach the surface within the time step. In this case, the particle is simply moved to its end of time step position
    \begin{equation}
        \pos_{p} = \pos_{0,p} + c\Delta t_p\direction.
    \end{equation}
    The remaining time of the particle is updated by
    \begin{equation}
        \Delta t'_p = \Delta t_p - \Delta t_{s,i}.
    \end{equation}
    
    If a particle crosses a surface that is part of the boundary, it is always reflected back into the domain. Therefore, the number of particles remain constant throughout the calculation. The new direction is found by the reflection law
    \begin{equation} \label{eq:method:ho:reflection}
        \direction_{p}' = \direction_{p} - 2\left(\normal_{j} \vd \direction_{p}\right) \cdot \normal_{j}
    \end{equation}
    and the intensity is
    \begin{equation}
        {I'}^\HO_{p,j,n} = \alpha_{j} I^\HO_{p,j,n} + \frac{ac T_{\mathrm{BC},j}^4}{4\pi}
    \end{equation}
    where \(\alpha_{j}\) is the reflection factor. This provides both vacuum (\(\alpha = 0\)) and reflective (\(\alpha = 1\)) conditions, and in between. For vacuum boundaries with no influx, the intensity is set to \({I'}^{\HO}_{p,j,n} = 0\), but it evolves according to \cref{eq:method:ho:characteristic}.

\subsubsection{Linear source approximation and tallying}
    To solve the characteristic equation, \cref{eq:method:ho:characteristic}, effectively, we must be able to evaluate the source term \(Q^{LO}\left(\pos_{p}\left(t\right), t\right)\) given LO quantities. 
	A key consideration for the source evaluation is the need to capture the asymptotic diffusion limit (a critical numerical property~\cite{larsen_asymptotic_1987,larsen_asymptotic_1989,larsen_asymptotic_1983}), for which a linear (or higher order) representation of the relevant quantities \(E\) and \(T\) is needed~\cite{ferrer_linear_2012,ferrer_linear_2016,ferrer_linear_2018,wollaeger_implicit_2016}. Here, we consider a linear source reconstruction. However, linear descriptions are not without issues. They increase the computational cost per cell and, depending on the slope, may violate positivity (see \cref{sec:appendix:negative_temperatures} for our treatment to enforce positivity of the source). This may occur for cells with low temperatures and steep gradients, e.g., at boundary layers and  at thermal fronts. Finally, not all linear source reconstructions capture correctly the asymptotic diffusion limit. In what follows, following~\citet{ferrer_linear_2016,ferrer_linear_2018}, we first outline a general linear-reconstruction procedure for an arbitrary function that will yield a method able to capture the asymptotic diffusion limit. Later, we use this reconstruction for the emission source in the HO solver, and derive the corresponding moment tallies needed.
	
\paragraph{Linear reconstruction procedure}
    
    Let \(\phi_{i,n}\left(\pos\right)\) be a quantity evaluated by scoring of particles in cell \(i\) and timestep \(n\), which can be an arbitrary function in \(\pos \in V_i\) . We seek its linear representation \(\psi_{i,n}\left(\pos\right)\) within the same cell \(i\). For simplicity, let \(\phi_{i,n}\) be angle-independent, to focus on the spatial aspect. Its zeroth spatial moment is
    \begin{align}\label{eq:method:ho:linear:phi_zeroth}
    \avg{\phi}_{i,n} &= \frac{1}{V_{i}}\int_{V_{i}} \phi_{i,n}\left(\pos\right) \dx[V] \notag \\
    &= \frac{1}{4\pi V_{i}} \sum_{p = 1}^{P_{i}} \weight[p] \int_{0}^{s_{p,i,n}} \phi_{i,n}\left(\pos\left(s'\right)\right) \dx[s'],
    \end{align}
    and its first spatial moment is
    \begin{align}\label{eq:method:ho:linear:phi_first}
    \linearv{\phi}_{i,n} &= \frac{1}{V_{i}}\int_{V_{i}} \pos \phi_{i,n}\left(\pos\right) \dx[V] \notag \\
    &= \frac{1}{4\pi V_{i} } \sum_{p = 1}^{P_{i}} \weight[p] \int_{0}^{s_{p,i,n}} \pos\left(s'\right) \phi_{i,n}\left(\pos\left(s'\right)\right) \dx[s'],
    \end{align}
    where \(s_{p,i,n}\) is the track of particle \(p\) in cell \(i\) at time step \(n\), \weight[p] denotes the particle's phase-space volume, and \(P_i\) is the number of particles in cell \(i\).
    
     We consider the following ansatz for the linear representation of \(\phi_{i,n}\) in cell \({i}\)
    \begin{equation}\label{eq:method:ho:linear:model}
        \psi_{i,n}\left(\pos\right) = \avg{\psi}_{i,n} + \linearv{\psi}_{i,n} \cdot \left(\pos - \ccenter\right)
    \end{equation}
    where \(\avg{\psi}_{i,n}\) is the constant part and the vector \(\linearv{\psi}_{i,n}\) is its gradient. We will adopt this notation also for other linear quantities throughout this paper. In \cref{eq:method:ho:linear:model},
    \begin{equation}
        \ccenter = \frac{1}{V_{i}} \int_{V_{i}} \pos \dx[V].
    \end{equation}
    is the center of mass of the cell \(i\). The following property follows:
    \begin{equation} \label{eq:method:ho:linear:psi_average}
        \frac{1}{V_{i}} \int_{V_{i}} \psi_{i,n}\left(\pos\right) \dx[V] = \avg{\psi}_{i,n}
    \end{equation}
    Also, by definition:
    \begin{equation}
	    \frac{1}{V_{i}}  \int_{V_i}\left(\pos - \ccenter\right) \dx[V] = 0.
    \end{equation}
	Since the spatial moments of the linear representation must equal the scored spatial moments, the zeroth moment must satisfy:
    \begin{equation}
        \frac{1}{V_{i}} \int_{V_{i}} \psi_{i,n}\left(\pos\right) \dx[V] = \avg{\phi}_{i,n}
    \end{equation}
    and hence with \cref{eq:method:ho:linear:psi_average}
    \begin{equation}
        \avg{\psi}_{i,n} = \avg{\phi}_{i,n}.
    \end{equation}
    To approximate the gradient, we compute the first spatial moment as:
    \begin{equation}
        \frac{1}{V_{i}} \int_{V_{i}} \left(\pos - \ccenter\right) \psi_{i,n}\left(\pos\right) \dx[V] = \frac{1}{V_{i}} \int_{V_{i}} \left(\pos - \ccenter\right) \phi_{i,n}\left(\pos\right) \dx[V],
    \end{equation}
    which with \cref{eq:method:ho:linear:phi_zeroth,eq:method:ho:linear:phi_first,eq:method:ho:linear:model} gives the equation system
    \begin{equation}
        \linearv{\psi}_{i,n} \frac{1}{V_{i}}\int_{V_{i}} \left(\pos - \ccenter\right) \otimes \left(\pos - \ccenter\right)   \dx[V]
        = \linearv{\phi}_{i,n} - \ccenter \avg{\phi}_{i,n}
    \end{equation}
    where \(\otimes\) denotes the tensor product. Written algebraically, the solution is
    \begin{equation} \label{eq:method:ho:linear:matrix_eq}
        \linearv{\psi}_{i,n} = \mat{M}_{i}^{-1}\left(\linearv{\phi}_{i,n} - \avg{\phi}_{i,n} \ccenter\right)
    \end{equation}
    where the matrix
    \begin{equation}
        \mat{M}_{i} = \frac{1}{V_{i}}\int_{V_{i}} \left(\pos - \ccenter\right) \otimes \left(\pos - \ccenter\right) \dx[V]
    \end{equation}
    only contains geometric information~\cite{steger_calculation_1996}, and can be precomputed for each cell.
    
    \paragraph{Linear reconstruction of the emission source}
    Using the linear approximation, \cref{eq:method:ho:linear:model}, the source in cell \(i\) for time step \(n\) has the form
    \begin{equation}
       Q^\LO_{i,n}\left(\pos\right) = \avg{Q}^\LO_{i,n} + \linearv{Q}^\LO_{i,n} \,\vd \left(\pos - \ccenter\right).
    \end{equation}
    where \(\avg{Q}^\LO_{i,n} \) is the average and \(\linearv{Q}^\LO_{i,n}\) the source gradient.
	We begin by introducing the auxiliary variable
        \begin{equation} \label{eq:method:temperature:theta}
        \Theta_{i,n}\left(\pos\right) = T_{i,n}^4\left(\pos\right)
        \end{equation}
        so that the source term can be written as
        \begin{align}
        Q^\LO_{i,n}\left(\pos\right) &= \opacity[i,n]ac\Theta_{i,n}\left(\pos\right) \notag \\
        &= \opacity[i,n]ac\left(\avg{\Theta}_{i,n} + \linearv{\Theta}_{i,n}\cdot\left(\pos - \ccenter\right)\right).
        \end{align}
        The temperature is linearized by expanding \(\Theta_{i,n}\left(\pos\right)\) using a Taylor series
        \begin{align} \label{eq:method:temperature:temp_linear}
        T_{i,n}\left(\pos\right) &= \sqrt[4]{\avg{\Theta}_{i,n}} + \frac{1}{4} \avg{\Theta}_{i,n}^{\,-\frac{3}{4}} \linearv{\Theta} \vd \left(\pos - \ccenter\right) \notag \\
        &= \avg{T}_{i,n} + \linearv{T}_{i,n} \vd \left(\pos - \ccenter\right).
        \end{align}
        The cell-average temperature, $\avg{T}_{i,n}$, is updated according to the evolution equation obtained by integrating \cref{eq:method:lo_system_consistent:temperature} over time step \(n\) and cell \(i\) [and using \cref{eq:method:ho:linear:psi_average}]:
        \begin{equation} \label{eq:method:temperature:average}
        \frac{\density\cv }{\Delta t_n} \left(\avg{T}_{i,n} - \avg{T}_{i,n-1}\right) + \opacity[i,n] ac\avg{\Theta}_{i,n} - \opacity[i,n] c\avg{E}^\LO_{i,n} = 0
        \end{equation}
        For the temperature gradient, $ \linearv{T}_{i,n}$, we use the first spatial moment of \cref{eq:method:lo_system_consistent:temperature},
        \begin{equation}
        \frac{1}{V_{i}\Delta t_n} \int_{\Delta t_{n}} \int_{V_i}\left(\pos - \ccenter\right) \left[\density\cv \ddt[T_{i,n}\left(\pos\right)] + \opacity ac\Theta_{i,n}\left(\pos\right) - \opacity[i,n] cE^\LO_{i,n}\left(\pos\right)\right] \dx[V] \dt = 0
        \end{equation}
        to obtain
        \begin{equation} \label{eq:method:linear:temperature_first_moment}
        \frac{1}{V_{i}} \int_{V_i}\left(\pos - \ccenter\right) \otimes \left(\pos - \ccenter\right) \dx[V]
        \cdot \left[\frac{\density\cv }{\Delta t_n} \left(\linearv{T}_{i,n} - \linearv{T}_{i,n-1}\right) + \opacity[i,n] ac\linearv{\Theta}_{i,n} - \opacity[i,n] c\linearv{ E}^\LO_{i,n}\right]
        = 0.
        \end{equation}
        Using \cref{eq:method:temperature:temp_linear}, we can solve this equation for the gradient of $\Theta_{i,n}$ as a function of the gradient of $E^{LO}_{i,n}$:
        \begin{equation}
        \linearv{\Theta}_{i,n} = \frac{\frac{\density\cv}{4\Delta t_{n}}\avg{\Theta}_{i,n-1}^{\,-\frac{3}{4}} \linearv{\Theta}_{i,n-1} + \opacity[i,n]c\linearv{E}^\LO_{i,n}} {\frac{\density\cv}{4\Delta t_{n}}\avg{\Theta}_{i,n}^{\,-\frac{3}{4}} + \opacity[i,n]ac}.
        \end{equation}
	    For simplicity, the LO gradient of $E_{i,n}$ is found in this study by scaling its HO gradient as:
	    \begin{equation}	
		    \linearv{E}^\LO_{i,n} = \linearv{E}^\HO_{i,n} \frac{\avg{E}^\LO_{i,n}}{\avg{E}^\HO_{i,n}}
	    \end{equation}
	    A better choice would be to discretize the LO system with Discontinuous Galerkin (DG), but we leave this for future work. Thus, all that remains is to tally the HO average and gradient components of $E_{i,n}$. We explain next how this is done.
    
    \paragraph{Tallying of HO moments}
	    Combining the linearized emission source, \cref{eq:method:temperature:temp_linear}, with the particle equation of motion, \cref{eq:method:ho:ray}, yields the emission source for a specific particle \(p\):
    \begin{equation}
        Q^\LO_{p,i,n}\left(s\right) = \avg{Q}^\LO_{i,n} + \linearv{Q}^\LO_{i,n} \vd \left(\pos_{0,p,i,n} - \ccenter + s\direction_{p}\right)  \notag \\
        = \avg{q}^\LO_{p,i,n} + s~ \linear{q}^\LO_{p,i,n},
    \end{equation}
    which is a linear function of the orbit distance, $s$. Within a cell, the material properties and opacities are assumed to be constant and given by
    \begin{equation}
        \opacity[i,n] = \opacity\left(\ccenter, \avg{T}_{i,n-1} \right).
    \end{equation}     
    We can now analytically solve the integral in \cref{eq:method:ho:characteristic} and obtain the intensity function
    \begin{equation} \label{eq:method:ho:intensity}
        I^\HO_{p,i,n}\left(s\right) = I^\HO_{p,i,n}\left(0\right) \e{-\opacity[i,n] s}
        +  \left(\avg{q}^\LO_{p,i,n} - \frac{\linear{q}^\LO_{p,i,n}}{\opacity[i,n]}\right)\frac{G\left(\opacity[i,n] s\right)}{\opacity[i,n]} +s \frac{\linear{q}^\LO_{p,i,n}}{\opacity[i,n]},
    \end{equation}
    with \(G\left(\tau\right) = \left(1 - \e{-\tau}\right)\). Note that, with the known emission source, the particle intensity asymptotes to the equilibrium solution (instead of zero) in optically thick regimes, which results in much improved behavior compared to MC. 
    
    With the intensity analytically known, we can analytically tally the contribution of particle \(p\) in cell \(i\) to the average and gradient of the radiation energy density. The average radiation energy density per particle is found as:
    \begin{align} \label{eq:method:ho:delta_E_avg}
        \avg{\delta E}^\HO_{p,i,n} 
        &=  \int_{0}^{s_{p,i,n}} I^\HO_{p,i,n}\left(s'\right) \dx[s'] \notag \\
        &= I^\HO_{p,i,n}\left(0\right)\frac{G\left(\opacity[i,n] s_{p,i,n}\right) }{\opacity[i,n]} +  \frac{1}{\opacity[i,n]}\left(\avg{q}^\LO_{p,i,n} - \frac{\linear{q}^\LO_{p,i,n}}{\opacity[i,n]}\right) \notag \\
        &\qquad\cdot \left(s_{p,i,n} - \frac{G\left(\opacity[i,n] s_{p,i,n}\right)}{\opacity[i,n]}\right) 
            + \frac{s_{p,i,n}^2}{2\opacity[i,n]} \linear{q}^\LO_{p,i,n}.
    \end{align}
    The average radiation energy density in cell \(i\) is found as the sum of all particle contributions
    \begin{equation} \label{eq:method:ho:E_avg}
        \avg{E}^\HO_{i,n} = \frac{1}{V_{i} c^2\Delta t_n}\sum_{p = 1}^{P_{i}} \weight[p] \avg{\delta E}^\HO_{p,i,n}.
    \end{equation}

    Per \cref{eq:method:ho:linear:phi_first}, the gradient of the radiation energy density is calculated from the first spatial moment of the intensity function, with a single particle contribution given by:
    \begin{align} \label{eq:method:ho:delta_E_linear}
        { \linearv{\delta E}}^\HO_{p,i,n} &= \int_{0}^{s_{p,i,n}} \pos\left(s'\right) I^\HO_{p,i,n}\left(s'\right) \dx[s'] 
        \notag \\
        &= \pos_{0,p,i,n} \int_{0}^{s_{p,i,n}}  I^\HO_{p,i,n}\left(s'\right) \dx[s'] 
        + \direction_{p} \int_{0}^{s_{p,i,n}} s' I^\HO_{p,i,n}\left(s'\right) \dx[s'],
    \end{align}
    where we used \cref{eq:method:ho:ray} for the particle orbit.
    The first integral in \cref{eq:method:ho:delta_E_linear} is \( \avg{\delta E}_{p,i,n}\), \cref{eq:method:ho:delta_E_avg}, and the second integral  gives
    \begin{multline}
         \int_{0}^{s_{p,i,n}} s' I^\HO_{p,i,n}\left(s'\right) \dx[s'] 
         =  \left(\frac{G\left(\opacity[i,n] s_{p,i,n}\right)}{\opacity[i,n]^2} -  \frac{s_{p,i,n}}{\opacity[i,n]}\e{-\opacity[_{i,n}] s_{p,i,n}} \right) I^\HO_{p,i,n}\left(0\right)
         + \frac{s_{p,i,n}^3}{3\opacity[i,n]}  \linear{q}^\LO_{p,i,n}
         \\
         + \left(\avg{q}^\LO_{p,i,n} - \frac{\linear{q}^\LO_{p,i,n}}{\opacity[i,n]}\right)\left(\frac{s_{p,i,n}^2}{2\opacity[i,n]} + \frac{s_{p,i,n}}{\opacity[i,n]^2}\e{-\opacity[i,n] s_{p,i,n}} - \frac{G\left(\opacity[i,n] s_{p,i,n}\right)}{\opacity[i,n]^3}\right).
    \end{multline}
    Projecting the gradient according to \cref{eq:method:ho:linear:matrix_eq}, the linear reconstruction of the gradient of the radiation energy density reads:
    \begin{equation} \label{eq:method:ho:E_linear}
        \linearv{E}^\HO_{i,n} = \mat{M}^{-1} \left[\frac{1}{V_{i} c^2\Delta t_{n}}\sum_{p=1}^{P_{i}} \weight[p]{\linearv{\delta E}}^\HO_{p,i,n} - \ccenter\avg{E}^\HO_{i,n}\right].
    \end{equation}
    
    It is useful to point out that, in voids (\(\opacity[i,n] = 0\)) \cref{eq:method:ho:intensity,eq:method:ho:delta_E_avg,eq:method:ho:delta_E_linear} simplify to
    \begin{subequations}
        \begin{align}
            I^\HO_{p,i,n}\left(s\right) &= I^\HO_{p,i,n} \left(0\right), \\
            \avg{\delta E}^\HO_{p,i,n} &= s I^\HO_{p,i,n} \left(0\right), \\
            \linearv{\delta E}^\HO_{p,i,n} &= \half[s^2] I^\HO_{p,i,n} \left(0\right) \direction_{p},
        \end{align}
    \end{subequations}
    while \cref{eq:method:ho:E_avg,eq:method:ho:E_linear} remain the same. 
    
    Other required tallies include the end-of-time-step radiation energy density, which is the census of particles within a cell:
    \begin{equation}
        \avg{E}^\HO_{i,n+\half} = \frac{1}{cV_{i}} \sum_{p=1}^{P_{i}} \weight[p] I^\HO_{p,i,n+\half},
    \end{equation}
    and the partial fluxes across surface \(j\), which are the sum of intensities of all particles crossing it:
    \begin{subequations}
        \begin{align}
            f^{+\HO}_{j,n} &= \frac{1}{A_j c\Delta t_n} \sum_{p=1}^{P_{j}} \weight[p] I^\HO_{p,j,n} \qquad \text{for} \quad \direction_{p} \vd \normal_{j} > 0,
            \\
            f^{-\HO}_{j,n} &= \frac{1}{A_j c\Delta t_n} \sum_{p=1}^{P_{j}} \weight[p] I^\HO_{p,j,n} \qquad \text{for} \quad \direction_{p} \vd \normal_{j} < 0.
        \end{align}
    \end{subequations}

\subsection{Multi-frequency extension} \label{sec:method:mf}
    For frequency-dependent problems, we employ the standard multi-frequency discretization. The specific angular intensity \(I_g\) for group \(g\) is defined as
    \begin{equation}
        I^\HO_g\left(\pos, \direction, t\right) = \int_{\frequency_{g - \half}}^{\frequency_{g+\half}} I\left(\pos, \direction, \frequency, t\right) \dx[\frequency].
    \end{equation}
    This leads to the multi-group TRT equation
     \begin{equation}
        \frac{1}{c} \ddt[I^\HO_g] + \direction \vd \del I^\HO_g + \opacity[g] I^\HO_g = \frac{\opacity[g] b_g ac T^4}{4\pi},
    \end{equation}
    where 
    \begin{equation}
        b_g\left(T\right) = \frac{\int_{\nu_{g - \half}}^{\nu_{g+\half}} B\left(\nu, T\right) \dx[\nu]}{\int_{0}^{\infty} B\left(\nu, T\right) \dx[\nu]} 
        = \frac{\int_{\nu_{g-\half}}^{\nu_{g+\half}} B\left(\nu, T\right) \dx[\nu]}{\frac{1}{4\pi}acT^4}
    \end{equation}
    is the Planck Spectrum factor. The characteristic solution along a particle trajectory, \cref{eq:method:ho:characteristic}, can be written as
    \begin{equation} \label{eq:method:mf:characteristic}
        I^\HO_{g,p}\left(t\right) = I^\HO_{g,p}\left(t_0\right) \e{-\int_{t_0}^{t} \opacity[g] c\dt'}
        + \int_{t_0}^{t} \e{-\int_{t'}^{t}\opacity[g] c\dt''} Q^\LO_g\left(\pos_{p}\left(t'\right), t'\right) c\dt'
    \end{equation}
    with
    \begin{equation}
        Q^\LO_g\left(\pos, t'\right) = \frac{b_g a c T^4\left(\pos, t\right)}{4\pi} .
    \end{equation}
    Using the same procedure as in \cref{sec:method:ho}, the group-wise quantities \(I^\HO_{p,i,n}\left(s\right)\), \(\avg{\delta E}^\HO_{g,p,i,n} \), \({\linearv{\delta E}}^\HO_{g,p,i,n}\)  are calculated the same as the gray counterparts in \cref{eq:method:ho:intensity,eq:method:ho:delta_E_avg,eq:method:ho:delta_E_linear}, respectively, using \(I^\HO_{g,p,i,n}\), \(\avg{q}^\LO_{g,p,i,n}\), \(\linear{q}^\LO_{g,p,i,n}\) and \opacity[g,i,n]. The tallied quantities then become
    \begin{align}
        \avg{E}^\HO_{i,n} &= \frac{1}{V_{i} c^2\Delta t_n}\sum_{p = 1}^{P_{i}} \weight[p] \sum_{g=1}^{G} \avg{\delta E}^\HO_{g,p,i,n} \\
        \linearv{E}^\HO_{i,n} &= \mat{M}^{-1} \left[\frac{1}{V_{i} c^2\Delta t_{n}}\sum_{p=1}^{P_{i}} \weight[p] \sum_{g=1}^{G} {\linearv{\delta E}}^\HO_{g,p,i,n} - \ccenter\avg{E}^\HO_{i,n}\right] \\
        \avg{E}^\HO_{i,n+\half} &= \frac{1}{cV_{i}} \sum_{p=1}^{P_{i}} \weight[p] \sum_{g=1}^{G} I^\HO_{g,p,i,n+\half} \\
        f^{\pm\HO}_{j,n} &= \frac{1}{A_j c\Delta t_n} \sum_{p=1}^{P_{j}} \weight[p] \sum_{g=1}^{G} I^\HO_{g,p,j,n} 
        \qquad \text{for} \quad \direction_{p} \vd \normal_{j} \gtrless 0
    \end{align}
    with the only difference being the summation over the groups. 
    
    The LO-system remains gray in the multi-frequency case. However, the opacities must be changed to weighted opacities~\cite{yee_stable_2017}. The LO-system becomes
    \begin{subequations}
        \begin{align}
            \ddt[E^\LO] + \div \vec{F}^\LO + \opacityE^\HO c E^\LO - \opacityP ac T^{4} &= \resSource, \\
            \frac{1}{c} \ddt[\vec{F}^\LO] +  \frac{c}{3}\del E^\LO + \opacityR \vec{F}^\LO &= \vec{\gamma}^\HO cE^\LO, \\
            \density \cv \ddt[T] + \opacityP ac T^4 -  \opacityE^\HO cE^\LO &= 0,
        \end{align}
    \end{subequations}
    where
    \begin{subequations}
        \begin{equation}
            \opacityE^\HO \equiv \frac{\int_{0}^{\infty} \opacity\int_{4\pi} I^\HO \domg  \dx[\frequency]}{\int_{0}^{\infty} \int_{4\pi} I^\HO \domg \dx[\frequency]} 
            =  \frac{\sum_{g=1}^{G} \opacity[g]\int_{4\pi} I^\HO_g\domg}{cE^\HO}
        \end{equation}
    is the radiation weighted opacity evaluated from the HO system,
        \begin{equation}
            \opacityP \equiv  \frac{\int_{0}^{\infty} \opacity B\left(\frequency, T\right) \dx[\frequency]}{\int_{0}^{\infty} B\left(\frequency, T\right) \dx[\frequency]} 
            = \sum_{g=1}^{G} \opacity[g] b_g
        \end{equation}
    \end{subequations}
    is the Planck weighted opacity, and 
    \begin{equation}
            \opacityR = \frac{\int_{0}^{\infty} \dd[B]{T}\big\vert_{T} \dx[\frequency]}
            {\int_{0}^{\infty} \frac{1}{\opacity\left(\frequency, T\right)}\dd[B]{T}\big\vert_{T} \dx[\frequency]}
            = \frac{\sum_{g=1}^{G} \dd[B_g]{T}\big\vert_{T}}{\sum_{g=1}^{G} \frac{1}{\opacity[g]}\dd[B_g]{T}\big\vert_{T}}
    \end{equation}
     is the Rosseland weighted opacity. It is important to note that we use the most up-to-date quantities \(E^\HO\), \(I^\HO_g\) and \(b_g\) to evaluate the weighted opacities, regardless the temporal centering of the multi-frequency opacities \opacity[g]. The solution strategy for the LO system remains the same as described for the gray case, \cref{sec:method:lo}.
    
    We emphasize here that there is a significant advantage of the DP method compared to IMC or \sn for multi-frequency problems in that the particle trajectory, \cref{eq:method:ho:ray}, is independent of frequency and thus all frequency information is carried by each particle. Thus, we can provide the same phase-space resolution with the same number of particles regardless of the number of frequency groups. Although each particle carries more information, and therefore the memory requirements will increase, we can use a single ray-tracing step per particle, thus reducing the computational effort per group.

\section{Numerical Results} \label{sec:results}

    In the following section, we present multi-dimensional numerical results for the Tophat and Hohlraum problems. One-dimensional demonstrations for the DP method can be found in earlier publications~\cite{hammer_multi-dimensional_2018,park_multigroup_2019}.

\subsection{Tophat problem}

    \begin{figure}
        \setlength{\figheight}{8cm}
        \begin{tikzpicture} 

\usetikzlibrary{patterns}

\def \scale{  1.  }
\def \xoffset{  0.5  }
\def \yoffset{  0.25  }

% The vertices A,B,C,D define the reference plan (vertical)
\coordinate (A) at ({0.0*\scale},{0.0*\scale}); % lower left corner
\coordinate (B) at ({0.5*\scale},{0.0*\scale}); % lower right corner
\coordinate (C) at ({2.0*\scale},{0.0*\scale}); % lower wall lower left corner
\coordinate (D) at ({0.5*\scale},{2.5*\scale}); % lower wall lower right corner
\coordinate (E) at ({1.5*\scale},{2.5*\scale}); % right wall lower left corner
\coordinate (F) at ({0.0*\scale},{3.0*\scale}); % right wall lower right corner
\coordinate (G) at ({1.0*\scale},{3.0*\scale}); % lower wall upper left corner
\coordinate (H) at ({0.0*\scale},{4.0*\scale}); % lower wall upper right corner
\coordinate (I) at ({1.0*\scale},{4.0*\scale}); % center lower left corner
\coordinate (J) at ({0.5*\scale},{4.5*\scale}); % center lower righ corner
\coordinate (K) at ({1.5*\scale},{4.5*\scale}); % center upper left corner
\coordinate (L) at ({0.0*\scale},{7.0*\scale}); % center upper right corner
\coordinate (M) at ({0.5*\scale},{7.0*\scale}); % center upper right corner
\coordinate (N) at ({2.0*\scale},{7.0*\scale}); % center upper right corner

% draw background
\draw[pattern=checkerboard light gray] (F) -- (G) -- (I) -- (H) -- cycle;
\draw[pattern=checkerboard light gray] (B) -- (C) -- (N) -- (M) -- (J) -- (K) -- (E) -- (D) -- cycle;

% Draw the horizontal edges
\draw[-,very thick] (A) --  (C)
                        (D) -- (E)
                        (F) -- (G)
                        (H) -- (I)
                        (J) -- (K)
                        (L) -- (N);

% Draw the vertical edges
\draw[-,very thick] (A) --  (L)
                              (B) -- (D)
                              (E) -- (K)
                              (G) -- (I)
                              (J) -- (M)
                              (C) -- (N);
                       
\coordinate (X1) at ({0.0*\scale},{0.25*\scale}); % upper right corner
\coordinate (X2) at ({0.0*\scale},{2.75*\scale}); % upper right corner
\coordinate (X3) at ({1.25*\scale},{3.5*\scale}); % upper right corner
\coordinate (X4) at ({0.0*\scale},{4.25*\scale}); % upper right corner
\coordinate (X5) at ({0.0*\scale},{6.75*\scale}); % upper right corner
\coordinate (X6) at ({1.475*\scale},{2.275*\scale}); % upper right corner

%% Marking the vertices (red)
\fill[red]  (X1) circle [radius=2pt];
\fill[red]  (X2) circle [radius=2pt];
\fill[red]  (X3) circle [radius=2pt];
\fill[red]  (X4) circle [radius=2pt];
\fill[red]  (X5) circle [radius=2pt];
\fill[red]  (X6) circle [radius=2pt];

%% Name the vertices (the names are not consistent
%%  with the node name, but it makes the programming easier)
\draw (X1) node [right]     {\footnotesize \(X_1\) \scriptsize \(\left(0, 0.25\right)\)}
            (X2) node [right]           {\footnotesize \(X_2\)\scriptsize\(\left(0, 2.75\right)\)}
            (X3)  node [above]       {\footnotesize \(X_3\)}
            (X3)  node [left=+3pt]  {\tiny\(\left(1.25,3.5\right)\)}
            (X4)  node [right]            {\footnotesize \(X_4\)\scriptsize\(\left(0, 4.25\right)\)}
            (X5)  node [right]       {\footnotesize \(X_5\) \scriptsize \(\left(0, 6.75\right)\)}
            (X6)  node [below]       {\footnotesize \(X_6\)}
            (X6)  node [below=+15pt,left=-18pt] {\scriptsize\(\left(1.48, 2.28\right)\)};

% measurment lines

\coordinate (Y1) at ({0.0*\scale},{-\xoffset});
\coordinate (Y2) at ({0.5*\scale},{-\xoffset});
\coordinate (Y3) at ({1.0*\scale},{-\xoffset});
\coordinate (Y4) at ({1.5*\scale},{-\xoffset});
\coordinate (Y5) at ({2.0*\scale},{-\xoffset});

\draw[<->] (Y1) -- (Y2);
\draw (Y1) -- (Y2) node [midway, above, sloped] (TextNode) {\footnotesize 0.5};
\draw[<->] (Y2) -- (Y3);
\draw (Y2) -- (Y3) node [midway, above, sloped] (TextNode) {\footnotesize 0.5};
\draw[<->] (Y3) -- (Y4);
\draw (Y3) -- (Y4) node [midway, above] (TextNode) {\footnotesize 0.5};
\draw[<->] (Y4) -- (Y5);
\draw (Y4) -- (Y5) node [midway, above] (TextNode) {\footnotesize 0.5};

\coordinate (Z1) at ({-\yoffset},{0.0 * \scale});
\coordinate (Z2) at ({-\yoffset},{2.5 * \scale});
\coordinate (Z3) at ({-\yoffset},{3.0 * \scale});
\coordinate (Z4) at ({-\yoffset},{4.0 * \scale});
\coordinate (Z5) at ({-\yoffset},{4.5 * \scale});
\coordinate (Z6) at ({-\yoffset},{7.0 * \scale});

\draw[<->] (Z1) -- (Z2);
\draw (Z1) -- (Z2) node [midway, above, sloped] (TextNode) {\footnotesize 2.5};
\draw[<->] (Z2) -- (Z3);
\draw (Z2) -- (Z3) node [midway, above, sloped] (TextNode) {\footnotesize 0.5};
\draw[<->] (Z3) -- (Z4);
\draw (Z3) -- (Z4) node [midway, above, sloped] (TextNode) {\footnotesize 1.0};
\draw[<->] (Z4) -- (Z5);
\draw (Z4) -- (Z5) node [midway, above, sloped] (TextNode) {\footnotesize 0.5};
\draw[<->] (Z5) -- (Z6);
\draw (Z5) -- (Z6) node [midway, above, sloped] (TextNode) {\footnotesize 2.5};

%\draw (A) -- (B) node [midway, above, sloped] (TextNode) {\footnotesize \SI{500}{\eV}};

\end{tikzpicture}
        \caption{Layout of the Tophat problem, all measurements in \si{\cm}. The filled areas are the dense material, the rest is optically thin.}
        \label{fig:results:tophat:design}
    \end{figure}
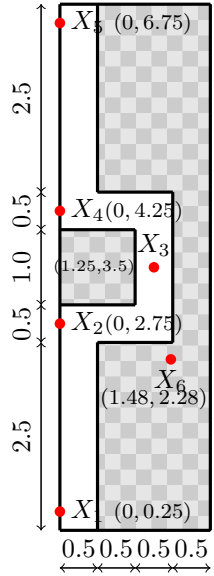

    The Tophat, or crooked pipe, problem is a two-dimensional problem in which a region of dense, opaque material (\(\density = \SI{10}{g\per\centi\metre\cubed}\), \(\cv = \SI{1e12}{erg\per\gram\per\electronvolt}\), \(\opacity = \SI{2000}{\per\centi\metre}\)) is embedded into a channel of thin material (\(\density = \SI{0.1}{g\per\centi\metre\cubed}\), \(\cv = \SI{1e12}{erg\per\gram\per\electronvolt}\), \(\opacity = \SI{0.2}{\per\cm}\)). The channel itself is surrounded by the opaque material. While the original definition was in cylindrical coordinates~\cite{gentile_implicit_2001}, we have adapted the problem to Cartesian coordinates. The problem is \SI{2}{\cm} by \SI{7}{\cm} with a reflective boundary condition at \(x = 0\), and vacuum on all other sides. \Cref{fig:results:tophat:design} shows the geometry of the problem with the corresponding measurements. The grayed regions contain the opaque material, while the white region is optically thin. The mesh used for the calculation is rectangular with \(\Delta x = \Delta y = \SI{0.05}{\cm}\), which gives \num{40} by \num{140} cells. In each cell, particles were initialized at the four centers of the corner subcells, using a \sn standard Gauss-Chebychev product quadrature with 8 polar and 24 azimuthal angles (see \cref{sec:appendix:particle_initalization} for details). This results in a total of \num{2150400} particles.
    
    At the beginning, the problem is in thermal equilibrium at the initial temperature \(T_0 = \SI{50}{\electronvolt}\) and a temperature source \(T_\mathrm{inc} = \SI{500}{\electronvolt}\) is applied to the bottom of the thin channel at \(y=\SI{0}{\centi\metre}\), \(x < \SI{0.5}{\centi\metre}\). The problem was run with an initial time step of \(\Delta t_0 = \SI{1e-12}{\second}\), which increased by a factor of 1.1 each step up to a maximum of \(\Delta t_\mathrm{max} = \SI{5e-11}{\second}\), for a total time of \(t_\mathrm{end} = \SI{1e-6}{\second}\). We used a tolerance of \(\tau_\mathrm{HOLO} = \SI{1e-4}{}\) for the HOLO solver with the convergence criteria
    \begin{equation}
        \frac{\norm[\infty]{\vec{E}^\HO_{n} - \vec{E}^\LO_{n}}}{\vec{E}^\LO_{n}} < \tau_\mathrm{HOLO},
    \end{equation}
    \(\tau_\mathrm{P1} = \SI{1e-8}{}\) for the P1 solver with the error calculated as the relative infinity norm of the nonlinear Newton-update, and \(\tau_T = \SI{1e-12}{}\) for the temperature solver with the relative Newton-update as error.
    
    With time, the radiation travels along the thin channel. The problem cannot be solved accurately with diffusion alone, since diffusion cannot model the flow of radiation around the corners. On the other hand, it is necessary for the algorithm to respect the asymptotic diffusion limit, or the radiation will diffuse too fast into the thick material.
    
    We use six points to track the temperature evolving over time, five in the thin material (\(X_1 - X_5\)), plus one in the thick material (\(X_6\)). The points and their corresponding coordinates are shown in \cref{fig:results:tophat:design}.

    \begin{figure}
        \begin{minipage}{0.45\textwidth}
            \includegraphics[width=\textwidth]{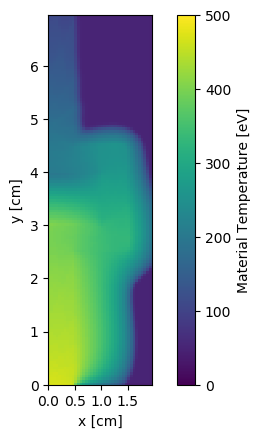}
            \subcaption{Flat source}
            \label{fig:results:tophat:square_temp:flat}
        \end{minipage} \hfill
        \begin{minipage}{0.45\textwidth}
            \includegraphics[width=\textwidth]{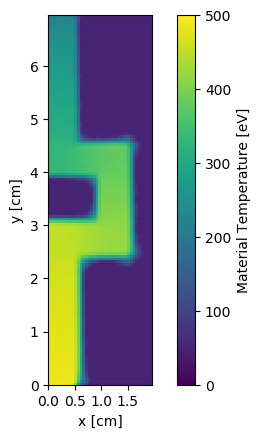}
            \subcaption{Linear source}
            \label{fig:results:tophat:square_temp:linear}
        \end{minipage}
        \caption{Material temperature for the Tophat problem at \SI{1e-6}{s} using a square mesh and flat- and linear-source approximations}
        \label{fig:results:tophat:square_temp}
    \end{figure}

    The material temperature at the final time \SI{1e-6}{\second} is shown in \cref{fig:results:tophat:square_temp}. These calculations use an orthogonal or square mesh. \cref{fig:results:tophat:square_temp} clearly shows the effect of the linear source approximation. Without a linear source, the radiation diffuses too fast into the thick material (\cref{fig:results:tophat:square_temp:flat}). Using the linear-source representation, the temperature only heats the first line of cells in the thick material except at the corners of the channel, giving a highly improved solution. The excessive numerical diffusion at the corners is caused by the lack of the bi-linear term in the source shape, which is necessary to recover the asymptotic diffusion limit on rectangular meshes~\cite{adams_characteristic_1998}. The error shows preferentially at the corners of the channel, because the linear representation cannot describe the temperature profile sufficiently there.

    \begin{figure}
        \setlength{\figheight}{8cm}
        \setlength{\figwidth}{0.45\textwidth}
        
        \begin{minipage}{0.45\textwidth}
            \begin{tikzpicture} 

\usetikzlibrary{patterns}

\def \scale{  1.  }
\def \xoffset{  0.5  }
\def \yoffset{  0.25  }

% The vertices A,B,C,D define the reference plan (vertical)
\coordinate (X00) at ({0*\scale},{0*\scale}); 
\coordinate (X10) at ({1*\scale},{0*\scale}); 
\coordinate (X20) at ({2*\scale},{0*\scale}); 
\coordinate (X30) at ({3*\scale},{0*\scale}); 
\coordinate (X40) at ({4*\scale},{0*\scale}); 

\coordinate (X01) at ({0*\scale},{1*\scale}); 
\coordinate (X11) at ({1*\scale},{1*\scale}); 
\coordinate (X21) at ({2*\scale},{1*\scale}); 
\coordinate (X31) at ({3*\scale},{1*\scale}); 
\coordinate (X41) at ({4*\scale},{1*\scale}); 

\coordinate (X02) at ({0*\scale},{2*\scale}); 
\coordinate (X12) at ({1*\scale},{2*\scale}); 
\coordinate (X22) at ({2*\scale},{2*\scale}); 
\coordinate (X32) at ({3*\scale},{2*\scale}); 
\coordinate (X42) at ({4*\scale},{2*\scale}); 

\coordinate (X03) at ({0*\scale},{3*\scale}); 
\coordinate (X13) at ({1*\scale},{3*\scale}); 
\coordinate (X23) at ({2*\scale},{3*\scale}); 
\coordinate (X33) at ({3*\scale},{3*\scale}); 
\coordinate (X43) at ({4*\scale},{3*\scale}); 

\coordinate (X04) at ({0*\scale},{4*\scale}); 
\coordinate (X14) at ({1*\scale},{4*\scale}); 
\coordinate (X24) at ({2*\scale},{4*\scale}); 
\coordinate (X34) at ({3*\scale},{4*\scale}); 
\coordinate (X44) at ({4*\scale},{4*\scale});

% Draw the horizontal edges
\draw[-,thick] (X00) --  (X10)
                        (X10) --  (X20)
                        (X20) --  (X30)
                        (X30) --  (X40)
                        
                        (X02) --  (X12)
                        (X12) --  (X22)
                        (X22) --  (X32)
                        (X32) --  (X42)
                        
                        (X04) --  (X14)
                        (X14) --  (X24)
                        (X24) --  (X34)
                        (X34) --  (X44)
                        
                        (X00) --  (X01)
                        (X01) --  (X02)
                        (X02) --  (X03)
                        (X03) --  (X04)
                        
                        (X20) --  (X21)
                        (X21) --  (X22)
                        (X22) --  (X23)
                        (X23) --  (X24)
                        
                        (X40) --  (X41)
                        (X41) --  (X42)
                        (X42) --  (X43)
                        (X43) --  (X44);

\draw[-,thick, dotted] 
                        (X01) --  (X11)
                        (X11) --  (X21)
                        (X21) --  (X31)
                        (X31) --  (X41)
                        
                        (X03) --  (X13)
                        (X13) --  (X23)
                        (X23) --  (X33)
                        (X33) --  (X43)

                      (X10) --  (X11)
                      (X11) --  (X12)
                      (X12) --  (X13)
                      (X13) --  (X14)
                      
                      (X30) --  (X31)
                      (X31) --  (X32)
                      (X32) --  (X33)
                      (X33) --  (X34);
                 
\coordinate (Y1) at ({0.0*\scale},{-\xoffset});
\coordinate (Y2) at ({1*\scale},{-\xoffset});
\coordinate (Y3) at ({2*\scale},{-\xoffset});
\coordinate (Y4) at ({3*\scale},{-\xoffset});
\coordinate (Y5) at ({4*\scale},{-\xoffset});

\draw[<->] (Y1) -- (Y2);
\draw[<->] (Y2) -- (Y3);
\draw[<->] (Y3) -- (Y4);
\draw[<->] (Y4) -- (Y5);

\draw (Y1) -- (Y2) node [midway, above, sloped] (TextNode) {\small \(\Delta x\)};
\draw (Y2) -- (Y3) node [midway, above, sloped] (TextNode) {\small \(\Delta x\)};
\draw (Y3) -- (Y4) node [midway, above, sloped] (TextNode) {\small \(\Delta x\)};
\draw (Y4) -- (Y5) node [midway, above, sloped] (TextNode) {\small \(\Delta x\)};

\coordinate (Z1) at ({-\yoffset},{0 * \scale});
\coordinate (Z2) at ({-\yoffset},{1 * \scale});
\coordinate (Z3) at ({-\yoffset},{2 * \scale});
\coordinate (Z4) at ({-\yoffset},{3 * \scale});
\coordinate (Z5) at ({-\yoffset},{4 * \scale});

\draw[<->] (Z1) -- (Z2);
\draw[<->] (Z2) -- (Z3);
\draw[<->] (Z3) -- (Z4);
\draw[<->] (Z4) -- (Z5);
\draw (Z1) -- (Z2) node [midway, above, sloped] (TextNode) {\small \(\Delta y\)};
\draw (Z2) -- (Z3) node [midway, above, sloped] (TextNode) {\small \(\Delta y\)};
\draw (Z3) -- (Z4) node [midway, above, sloped] (TextNode) {\small \(\Delta y\)};
\draw (Z4) -- (Z5) node [midway, above, sloped] (TextNode) {\small \(\Delta y\)};
                       
\end{tikzpicture}
            \subcaption{square-mesh}
        \end{minipage}
        \begin{minipage}{0.45\textwidth}
            \begin{tikzpicture} 

\usetikzlibrary{patterns}

\def \scale{  1.  }
\def \xoffset{  0.5  }
\def \yoffset{  0.25  }

% The vertices A,B,C,D define the reference plan (vertical)
\coordinate (X00) at ({0*\scale},{0*\scale}); 
\coordinate (X10) at ({1*\scale},{0*\scale}); 
\coordinate (X20) at ({2*\scale},{0*\scale}); 
\coordinate (X30) at ({3*\scale},{0*\scale}); 
\coordinate (X40) at ({4*\scale},{0*\scale}); 

\coordinate (X01) at ({0*\scale},{1*\scale}); 
\coordinate (X11) at ({1*\scale},{1*\scale}); 
\coordinate (X21) at ({2*\scale},{1*\scale}); 
\coordinate (X31) at ({3*\scale},{1*\scale}); 
\coordinate (X41) at ({4*\scale},{1*\scale}); 

\coordinate (X02) at ({0*\scale},{2*\scale}); 
\coordinate (X12) at ({1*\scale},{2*\scale}); 
\coordinate (X22) at ({2*\scale},{2*\scale}); 
\coordinate (X32) at ({3*\scale},{2*\scale}); 
\coordinate (X42) at ({4*\scale},{2*\scale}); 

\coordinate (X03) at ({0*\scale},{3*\scale}); 
\coordinate (X13) at ({1*\scale},{3*\scale}); 
\coordinate (X23) at ({2*\scale},{3*\scale}); 
\coordinate (X33) at ({3*\scale},{3*\scale}); 
\coordinate (X43) at ({4*\scale},{3*\scale}); 

\coordinate (X04) at ({0*\scale},{4*\scale}); 
\coordinate (X14) at ({1*\scale},{4*\scale}); 
\coordinate (X24) at ({2*\scale},{4*\scale}); 
\coordinate (X34) at ({3*\scale},{4*\scale}); 
\coordinate (X44) at ({4*\scale},{4*\scale});

% Draw the horizontal edges
\draw[-,thick] (X00) --  (X10)
                        (X10) --  (X20)
                        (X20) --  (X30)
                        (X30) --  (X40)
                        
                        (X02) --  (X12)
                        (X12) --  (X22)
                        (X22) --  (X32)
                        (X32) --  (X42)
                        
                        (X04) --  (X14)
                        (X14) --  (X24)
                        (X24) --  (X34)
                        (X34) --  (X44)
                        
                        (X00) --  (X01)
                        (X01) --  (X02)
                        (X02) --  (X03)
                        (X03) --  (X04)
                        
                        (X20) --  (X21)
                        (X21) --  (X22)
                        (X22) --  (X23)
                        (X23) --  (X24)
                        
                        (X40) --  (X41)
                        (X41) --  (X42)
                        (X42) --  (X43)
                        (X43) --  (X44);
                        
% triangles
\draw[-,thick,dashed] (X00) --  (X11) 
                                  (X20) --  (X31)
                                  (X11) --  (X22)
                                  (X31) --  (X42)
                                  (X02) --  (X13)
                                  (X22) --  (X33)
                                  (X13) --  (X24)
                                  (X33) --  (X44)
                                  
                                  (X11) --  (X20)
                                  (X31) --  (X40)
                                  (X02) --  (X11)
                                  (X22) --  (X31)
                                  (X13) --  (X22)
                                  (X33) --  (X42)
                                  (X04) --  (X13)
                                  (X24) --  (X33);   
                                  
\coordinate (Y1) at ({0.0*\scale},{-\xoffset});
\coordinate (Y2) at ({1*\scale},{-\xoffset});
\coordinate (Y3) at ({2*\scale},{-\xoffset});
\coordinate (Y4) at ({3*\scale},{-\xoffset});
\coordinate (Y5) at ({4*\scale},{-\xoffset});

\draw[<->] (Y1) -- (Y2);
\draw[<->] (Y2) -- (Y3);
\draw[<->] (Y3) -- (Y4);
\draw[<->] (Y4) -- (Y5);

\draw (Y1) -- (Y2) node [midway, above, sloped] (TextNode) {\small \(\Delta x\)};
\draw (Y2) -- (Y3) node [midway, above, sloped] (TextNode) {\small \(\Delta x\)};
\draw (Y3) -- (Y4) node [midway, above, sloped] (TextNode) {\small \(\Delta x\)};
\draw (Y4) -- (Y5) node [midway, above, sloped] (TextNode) {\small \(\Delta x\)};

\coordinate (Z1) at ({-\yoffset},{0 * \scale});
\coordinate (Z2) at ({-\yoffset},{1 * \scale});
\coordinate (Z3) at ({-\yoffset},{2 * \scale});
\coordinate (Z4) at ({-\yoffset},{3 * \scale});
\coordinate (Z5) at ({-\yoffset},{4 * \scale});

\draw[<->] (Z1) -- (Z2);
\draw[<->] (Z2) -- (Z3);
\draw[<->] (Z3) -- (Z4);
\draw[<->] (Z4) -- (Z5);
\draw (Z1) -- (Z2) node [midway, above, sloped] (TextNode) {\small \(\Delta y\)};
\draw (Z2) -- (Z3) node [midway, above, sloped] (TextNode) {\small \(\Delta y\)};
\draw (Z3) -- (Z4) node [midway, above, sloped] (TextNode) {\small \(\Delta y\)};
\draw (Z4) -- (Z5) node [midway, above, sloped] (TextNode) {\small \(\Delta y\)};
                       
\end{tikzpicture}
            \subcaption{Triangular mesh}
        \end{minipage}
        \caption{Comparison of part of the square and triangular mesh for the Tophat problem. Dotted lines delimit cells for the square mesh, while dashed delimit cells for the triangular mesh}
        \label{fig:results:tophat:mesh}
    \end{figure}
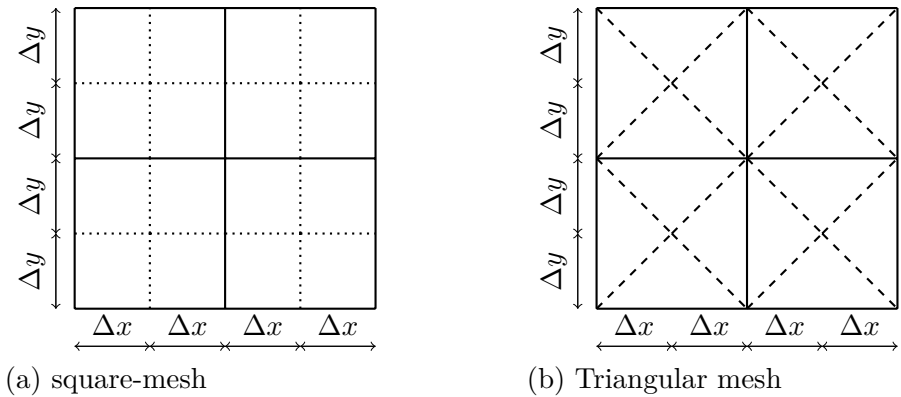

    \begin{figure}
        \begin{minipage}{0.45\textwidth}
            \includegraphics[width=\textwidth]{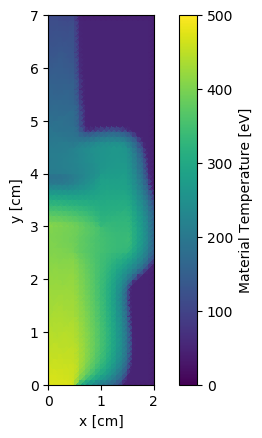}
            \subcaption{Flat source}
            \label{fig:results:tophat:tri_temp:flat}
        \end{minipage} \hfill
        \begin{minipage}{0.45\textwidth}
            \includegraphics[width=\textwidth]{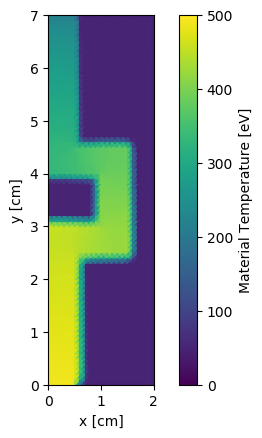}
            \subcaption{Linear source}
            \label{fig:results:tophat:tri_temp:linear}
        \end{minipage}
        \caption{Material temperature for the Tophat problem at \SI{1e-6}{s} using a triangular sub-mesh and flat- and linear-source approximations}
        \label{fig:results:tophat:tri_temp}
    \end{figure}

    The linear-source approximation is sufficient for the asymptotic diffusion limit on triangular meshes. To show this, we used a triangular mesh where four cells of the square mesh were combined and then divided into triangles using the center of these four cells as shown in \cref{fig:results:tophat:mesh}. This approach maintained the number of cells, particles and the area per cell, without introducing any preferred directionality. The results for the flat source do not show any significant change compared to the square mesh (\cref{fig:results:tophat:tri_temp:flat}), however the linear case does not exhibit the excessive numerical diffusion at the corner of the channel (\cref{fig:results:tophat:tri_temp:linear}). This demonstrates that the corner problem is caused by a lack of preservation of the asymptotic limit on the quadrilateral elements. 

    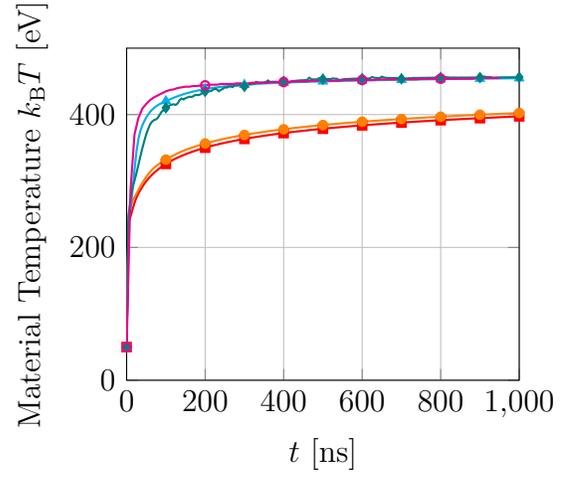
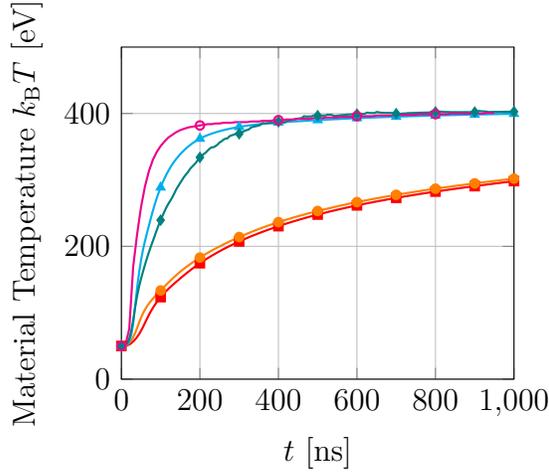
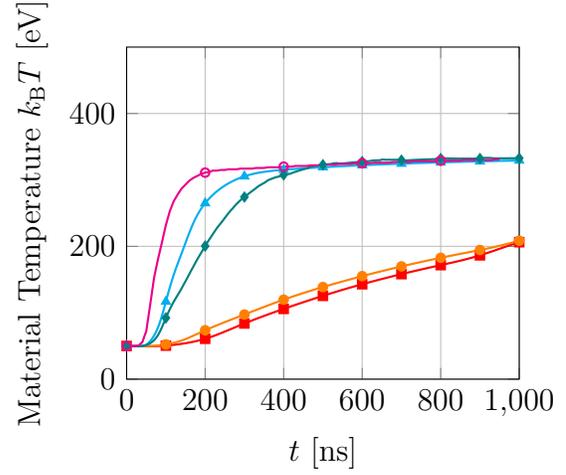
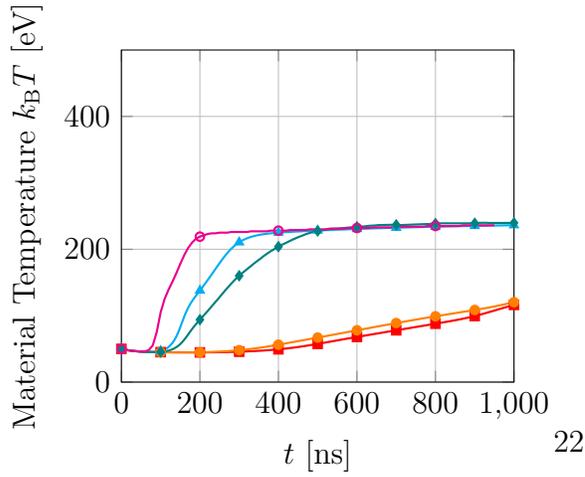
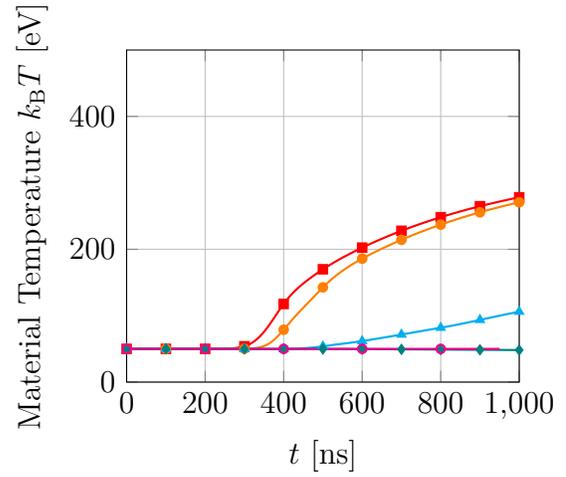
\begin{figure}
        \setlength{\figheight}{6cm}
        \setlength{\figwidth}{0.45\textwidth}
        
        \begin{minipage}{0.45\textwidth}
            \begin{tikzpicture}
    \begin{axis}[
            xmin = 0, xmax = 1000,
            ymin = 0, ymax = 500,
            y tick label style={/pgf/number format/.cd, scaled y ticks = false, fixed},
            x tick label style={/pgf/number format/.cd, scaled x ticks = false, fixed},
            height=\figheight,
            width=\figwidth,
            grid=major,
            legend pos = south east,
            legend columns = 1,
            xlabel = {\(t\) [\si{ns}]},
            ylabel = {Material Temperature \(\boltzmann T\) [\si{\electronvolt}]},
            mark options={solid, scale = 0.8, mark repeat = 20},
        ]

        \addplot [red, thick, mark=square*] table [x expr=\thisrow{t} * 1e9, y = x1_mat, col sep=comma] {data/tophat/tophat_flat_square.csv};
        \addlegendentry{flat, square}
        \addplot [orange, thick, mark=*] table [x expr=\thisrow{t} * 1e9, y = x1_mat, col sep=comma] {data/tophat/tophat_flat.csv};
        \addlegendentry{flat, tri}
        \addplot [cyan, thick, mark=triangle*] table [x expr=\thisrow{t} * 1e9, y = x1_mat, col sep=comma] {data/tophat/tophat_linear_square.csv};
        \addlegendentry{linear, square}
        \addplot [teal, thick, mark=diamond*] table [x expr=\thisrow{t} * 1e9, y = x1_mat, col sep=comma] {data/tophat/tophat_linear.csv};
        \addlegendentry{linear, tri}
        \addplot [magenta, thick, mark=o] table [x expr=\thisrow{t} * 1e9, y = x1_mat, col sep=comma] {data/tophat/comparison.csv};
        \addlegendentry{Capsaicin}
        
    \end{axis}
\end{tikzpicture}
            \subcaption{Point \(X_1\)}
            \label{fig:results:tophat:time_material:x1}
        \end{minipage} \hfill
        \begin{minipage}{0.45\textwidth}
            \begin{tikzpicture}
    \begin{axis}[
            xmin = 0, xmax = 1000,
            ymin = 0, ymax = 500,
            y tick label style={/pgf/number format/.cd, scaled y ticks = false, fixed},
            x tick label style={/pgf/number format/.cd, scaled x ticks = false, fixed},
            height=\figheight,
            width=\figwidth,
            grid=major,
            legend pos = south east,
            legend columns = 1,
            xlabel = {\(t\) [\si{ns}]},
            ylabel = {Material Temperature \(\boltzmann T\) [\si{\electronvolt}]},
            mark options={solid, scale = 0.8, mark repeat = 20},
        ]

        \addplot [red, thick, mark=square*] table [x expr=\thisrow{t} * 1e9, y = x2_mat, col sep=comma] {data/tophat/tophat_flat_square.csv};
        \addlegendentry{flat, square}
        \addplot [orange, thick, mark=*] table [x expr=\thisrow{t} * 1e9, y = x2_mat, col sep=comma] {data/tophat/tophat_flat.csv};
        \addlegendentry{flat, tri}
        \addplot [cyan, thick, mark=triangle*] table [x expr=\thisrow{t} * 1e9, y = x2_mat, col sep=comma] {data/tophat/tophat_linear_square.csv};
        \addlegendentry{linear, square}
        \addplot [teal, thick, mark=diamond*] table [x expr=\thisrow{t} * 1e9, y = x2_mat, col sep=comma] {data/tophat/tophat_linear.csv};
        \addlegendentry{linear, tri}
        \addplot [magenta, thick, mark=o] table [x expr=\thisrow{t} * 1e9, y = x2_mat, col sep=comma] {data/tophat/comparison.csv};
        \addlegendentry{Capsaicin}
        
        % hide legend
        \legend{}
    \end{axis}
\end{tikzpicture}
            \subcaption{Point \(X_2\)}
            \label{fig:results:tophat:time_material:x2}
        \end{minipage}
        \begin{minipage}{0.45\textwidth}
            \begin{tikzpicture}
    \begin{axis}[
            xmin = 0, xmax = 1000,
            ymin = 0, ymax = 500,
            y tick label style={/pgf/number format/.cd, scaled y ticks = false, fixed},
            x tick label style={/pgf/number format/.cd, scaled x ticks = false, fixed},
            height=\figheight,
            width=\figwidth,
            grid=major,
            legend pos = south east,
            legend columns = 1,
            xlabel = {\(t\) [\si{ns}]},
            ylabel = {Material Temperature \(\boltzmann T\) [\si{\electronvolt}]},
            mark options={solid, scale = 0.8, mark repeat = 20},
        ]

        \addplot [red, thick, mark=square*] table [x expr=\thisrow{t} * 1e9, y = x3_mat, col sep=comma] {data/tophat/tophat_flat_square.csv};
        \addlegendentry{flat, square}
        \addplot [orange, thick, mark=*] table [x expr=\thisrow{t} * 1e9, y = x3_mat, col sep=comma] {data/tophat/tophat_flat.csv};
        \addlegendentry{flat, tri}
        \addplot [cyan, thick, mark=triangle*] table [x expr=\thisrow{t} * 1e9, y = x3_mat, col sep=comma] {data/tophat/tophat_linear_square.csv};
        \addlegendentry{linear, square}
        \addplot [teal, thick, mark=diamond*] table [x expr=\thisrow{t} * 1e9, y = x3_mat, col sep=comma] {data/tophat/tophat_linear.csv};
        \addlegendentry{linear, tri}
        \addplot [magenta, thick, mark=o] table [x expr=\thisrow{t} * 1e9, y = x3_mat, col sep=comma] {data/tophat/comparison.csv};
        \addlegendentry{Capsaicin}
        
        % hide legend
        \legend{}
    \end{axis}
\end{tikzpicture}
            \subcaption{Point \(X_3\)}
            \label{fig:results:tophat:time_material:x3}
        \end{minipage} \hfill
        \begin{minipage}{0.45\textwidth}
            \begin{tikzpicture}
    \begin{axis}[
            xmin = 0, xmax = 1000,
            ymin = 0, ymax = 500,
            y tick label style={/pgf/number format/.cd, scaled y ticks = false, fixed},
            x tick label style={/pgf/number format/.cd, scaled x ticks = false, fixed},
            height=\figheight,
            width=\figwidth,
            grid=major,
            legend pos = north west,
            legend columns = 2,
            xlabel = {\(t\) [\si{ns}]},
            ylabel = {Material Temperature \(\boltzmann T\) [\si{\electronvolt}]},
            mark options={solid, scale = 0.8, mark repeat = 20},
        ]

        \addplot [red, thick, mark=square*] table [x expr=\thisrow{t} * 1e9, y = x4_mat, col sep=comma] {data/tophat/tophat_flat_square.csv};
        \addlegendentry{flat, square}
        \addplot [orange, thick, mark=*] table [x expr=\thisrow{t} * 1e9, y = x4_mat, col sep=comma] {data/tophat/tophat_flat.csv};
        \addlegendentry{flat, tri}
        \addplot [cyan, thick, mark=triangle*] table [x expr=\thisrow{t} * 1e9, y = x4_mat, col sep=comma] {data/tophat/tophat_linear_square.csv};
        \addlegendentry{linear, square}
        \addplot [teal, thick, mark=diamond*] table [x expr=\thisrow{t} * 1e9, y = x4_mat, col sep=comma] {data/tophat/tophat_linear.csv};
        \addlegendentry{linear, tri}
        \addplot [magenta, thick, mark=o] table [x expr=\thisrow{t} * 1e9, y = x4_mat, col sep=comma] {data/tophat/comparison.csv};
        \addlegendentry{Capsaicin}
        
        % hide legend
        \legend{}
    \end{axis}
\end{tikzpicture}
            \subcaption{Point \(X_4\)}
            \label{fig:results:tophat:time_material:x4}
        \end{minipage}
        \begin{minipage}{0.45\textwidth}
            \begin{tikzpicture}
    \begin{axis}[
            xmin = 0, xmax = 1000,
            ymin = 0, ymax = 500,
            y tick label style={/pgf/number format/.cd, scaled y ticks = false, fixed},
            x tick label style={/pgf/number format/.cd, scaled x ticks = false, fixed},
            height=\figheight,
            width=\figwidth,
            grid=major,
            legend pos = north west,
            legend columns = 1,
            xlabel = {\(t\) [\si{ns}]},
            ylabel = {Material Temperature \(\boltzmann T\) [\si{\electronvolt}]},
            mark options={solid, scale = 0.8, mark repeat = 20},
        ]

        \addplot [red, thick, mark=square*] table [x expr=\thisrow{t} * 1e9, y = x5_mat, col sep=comma] {data/tophat/tophat_flat_square.csv};
        \addlegendentry{flat, square}
        \addplot [orange, thick, mark=*] table [x expr=\thisrow{t} * 1e9, y = x5_mat, col sep=comma] {data/tophat/tophat_flat.csv};
        \addlegendentry{flat, tri}
        \addplot [cyan, thick, mark=triangle*] table [x expr=\thisrow{t} * 1e9, y = x5_mat, col sep=comma] {data/tophat/tophat_linear_square.csv};
        \addlegendentry{linear, square}
        \addplot [teal, thick, mark=diamond*] table [x expr=\thisrow{t} * 1e9, y = x5_mat, col sep=comma] {data/tophat/tophat_linear.csv};
        \addlegendentry{linear, tri}
        \addplot [magenta, thick, mark=o] table [x expr=\thisrow{t} * 1e9, y = x5_mat, col sep=comma] {data/tophat/comparison.csv};
        \addlegendentry{Capsaicin}
        
        % hide legend
        \legend{}
    \end{axis}
\end{tikzpicture}
            \subcaption{Point \(X_5\)} 
            \label{fig:results:tophat:time_material:x5}
        \end{minipage} \hfill
        \begin{minipage}{0.45\textwidth}
            \begin{tikzpicture}
    \begin{axis}[
            xmin = 0, xmax = 1000,
            ymin = 0, ymax = 500,
            y tick label style={/pgf/number format/.cd, scaled y ticks = false, fixed},
            x tick label style={/pgf/number format/.cd, scaled x ticks = false, fixed},
            height=\figheight,
            width=\figwidth,
            grid=major,
            legend pos = north west,
            legend columns = 1,
            xlabel = {\(t\) [\si{ns}]},
            ylabel = {Material Temperature \(\boltzmann T\) [\si{\electronvolt}]},
            mark options={solid, scale = 0.8, mark repeat = 20},
        ]

        \addplot [red, thick, mark=square*] table [x expr=\thisrow{t} * 1e9, y = x6_mat, col sep=comma] {data/tophat/tophat_flat_square.csv};
        \addlegendentry{flat, square}
        \addplot [orange, thick, mark=*] table [x expr=\thisrow{t} * 1e9, y = x6_mat, col sep=comma] {data/tophat/tophat_flat.csv};
        \addlegendentry{flat, tri}
        \addplot [cyan, thick, mark=triangle*] table [x expr=\thisrow{t} * 1e9, y = x6_mat, col sep=comma] {data/tophat/tophat_linear_square.csv};
        \addlegendentry{linear, square}
        \addplot [teal, thick, mark=diamond*] table [x expr=\thisrow{t} * 1e9, y = x6_mat, col sep=comma] {data/tophat/tophat_linear.csv};
        \addlegendentry{linear, tri}
        \addplot [magenta, thick, mark=o] table [x expr=\thisrow{t} * 1e9, y = x6_mat, col sep=comma] {data/tophat/comparison.csv};
        \addlegendentry{Capsaicin}
        
        % hide legend
        \legend{}
    \end{axis}
\end{tikzpicture}
            \subcaption{Point \(X_6\)}
            \label{fig:results:tophat:time_material:x6}
        \end{minipage}
        \caption{Material temperature over time for the six tracking points in the Tophat problem.}
        \label{fig:results:tophat:time_material}
    \end{figure}

    The time-dependent material temperature for the tracking points is shown in \cref{fig:results:tophat:time_material}. The further down the channel the tracking point is, the larger is the difference between the flat- and linear-source results (\crefrange{fig:results:tophat:time_material:x1}{fig:results:tophat:time_material:x5}), with the flat-source cases showing a much slower increase of temperature. The linear-source cases show also differences between the triangular and the square mesh in regard to how fast the heating occurs. While the square mesh shows faster heating, the final temperature is approximately the same as for the triangular mesh once an equilibrium is reached.
    
    The results for \(X_6\) (\cref{fig:results:tophat:time_material:x6}) show the material temperature away from the channel at a corner. It confirms the previous finding with regard to the diffusion limit. The cases using a flat-source approximation show a strong increase of the material temperature. The linear-source case on the square mesh also shows an increase in the temperature for later times, but it is significantly smaller than for the flat-source cases. The linear-source case on the triangular mesh, preserving the asymptotic diffusion limit, shows no increase at all for point \(X_6\).
    
    We have compared our results to Capsaicin~\cite{thompson_capsaicin:_2006}. Capsaicin shows faster transients than our code with the linear source, especially for points \(X_4\) and \(X_5\). But both codes show the same asymptotic solution for later times. The differences in the transients arise from the mesh, which is not sufficiently refined to resolve the boundary layer at the interface between the channel and the thick material. 
    
\FloatBarrier
\subsection{Hohlraum problem}

    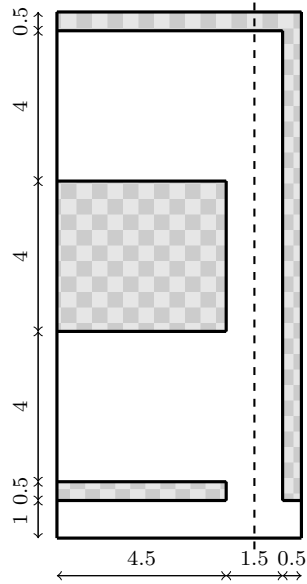
\begin{figure}
        \setlength{\figheight}{8cm}
        \begin{tikzpicture} 

\usetikzlibrary{patterns}

\def \scale{   5  }
\def \xoffset{  0.5  }
\def \yoffset{  0.25  }

% The vertices A,B,C,D define the reference plan (vertical)
\coordinate (A) at ({0*\scale},{0*\scale}); % lower left corner
\coordinate (B) at ({0.65*\scale},{0*\scale}); % lower right corner
\coordinate (C) at ({0*\scale},{0.10*\scale}); % lower wall lower left corner
\coordinate (D) at ({0.45*\scale},{0.1*\scale}); % lower wall lower right corner
\coordinate (E) at ({0.60*\scale},{0.10*\scale}); % right wall lower left corner
\coordinate (F) at ({0.65*\scale},{0.10*\scale}); % right wall lower right corner
\coordinate (G) at ({0*\scale},{0.15*\scale}); % lower wall upper left corner
\coordinate (H) at ({0.45*\scale},{0.15*\scale}); % lower wall upper right corner
\coordinate (I) at ({0*\scale},{0.55*\scale}); % center lower left corner
\coordinate (J) at ({0.45*\scale},{0.55*\scale}); % center lower righ corner
\coordinate (K) at ({0*\scale},{0.95*\scale}); % center upper left corner
\coordinate (L) at ({0.45*\scale},{0.95*\scale}); % center upper right corner
\coordinate (M) at ({0*\scale},{1.35*\scale}); % upper wall lower left corner
\coordinate (N) at ({0.6*\scale},{1.35*\scale}); % upper wall lower right corner
\coordinate (O) at ({0*\scale},{1.40*\scale}); % upper left corner
\coordinate (P) at ({0.65*\scale},{1.40*\scale}); % upper right corner

% draw background
\draw[pattern=checkerboard light gray] (C) -- (D) -- (H) -- (G) -- cycle;
\draw[pattern=checkerboard light gray] (I) -- (J) -- (L) -- (K) -- cycle;
\draw[pattern=checkerboard light gray] (E) --(F) -- (P) -- (O) -- (M) -- (N) -- cycle;

% Draw the horizontal edges
\draw[-,very thick] (A) --  (B)
                        (C) --  (D)
                        (E) -- (F)
                        (G)  --  (H)
                        (I) --  (J)
                        (K)  -- (L)
                        (M) -- (N)
                        (O) -- (P);

% Draw the vertical edges
\draw[-,very thick] (A) --  (O)
                       (D) -- (H)
                       (J) -- (L)
                       (E) -- (N)
                       (B) -- (P);
                       
\coordinate (X1) at ({0.225*\scale},{0.1*\scale}); % upper right corner
\coordinate (X2) at ({0.075*\scale},{0.55*\scale}); % upper right corner
\coordinate (X3) at ({0.375*\scale},{0.55*\scale}); % upper right corner
\coordinate (X4) at ({0.225*\scale},{0.95*\scale}); % upper right corner
\coordinate (X5) at ({0.6*\scale},{0.725*\scale}); % upper right corner
\coordinate (X6) at ({0.225*\scale},{1.35*\scale}); % upper right corner
\coordinate (X7) at ({0.525*\scale},{1.35*\scale}); % upper right corner
                       
%% Marking the vertices (red)
%\fill[red]  (X1) circle [radius=2pt];
%\fill[red]  (X2) circle [radius=2pt];
%\fill[red]  (X3) circle [radius=2pt];
%\fill[red]  (X4) circle [radius=2pt];
%\fill[red]  (X5) circle [radius=2pt];
%\fill[red]  (X6) circle [radius=2pt];
%\fill[red]  (X7) circle [radius=2pt];
%
%%% Name the vertices (the names are not consistent
%%%  with the node name, but it makes the programming easier)
%\draw (X1) node [below=+1pt]          {\footnotesize \(X_1\) \scriptsize\(\left(2.25, 1\right)\)}
%      (X2) node [below]                     {\footnotesize \(X_2\)}
%      (X2)  node [below=+15pt,left=-27.5pt] {\scriptsize\(\left(0.75, 5.5\right)\)}
%      (X3)  node [below]       {\footnotesize \(X_3\)}
%      (X3)  node [below=+15pt,left=-30] {\scriptsize\(\left(3.75,5.5\right)\)}
%      (X4)  node [above=+1pt]               {\footnotesize \(X_4\) \scriptsize\(\left(2.25, 9.5\right)\)}
%      (X5)  node [left=+1pt]                   {\footnotesize \(X_5\)}
%      (X5)  node [left=+11pt,below=+1pt]                {\tiny\(\left(6, 7.75\right)\)}
%      (X6)  node [below]                       {\footnotesize \(X_6\)}
%      (X6)  node [below=+15pt,left=-15pt] {\scriptsize\(\left(2.25,13.5\right)\)}
%      (X7) node [below]                         {\footnotesize \(X_7\)}
%      (X7)  node [below=+15pt, left=-13pt] {\scriptsize\(\left(5.25,13.5\right)\)};

% measurment lines

% wave front line

\coordinate (W1) at ({0.525*\scale},{-0.03*\scale});
\coordinate (W2) at ({0.525*\scale},{1.445*\scale});

\draw[-,thick,dashed] (W1) --  (W2);

\coordinate (Y1) at ({0.0*\scale},{-\xoffset});
\coordinate (Y2) at ({0.45*\scale},{-\xoffset});
\coordinate (Y3) at ({0.6*\scale},{-\xoffset});
\coordinate (Y4) at ({0.65*\scale},{-\xoffset});

\draw (Y1) -- (Y2) node [midway, above, sloped] (TextNode) {\scriptsize 4.5};
\draw (Y2) -- (Y3) node [midway, above, sloped] (TextNode) {\scriptsize 1.5};
\draw (Y3) -- (Y4) node [midway, above] (TextNode) {\scriptsize 0.5};

\draw[<->] (Y1) -- (Y2);
\draw[<->] (Y2) -- (Y3);
\draw[<->] (Y3) -- (Y4);

\coordinate (Z1) at ({-\yoffset},{0.0 * \scale});
\coordinate (Z2) at ({-\yoffset},{0.1 * \scale});
\coordinate (Z3) at ({-\yoffset},{0.15 * \scale});
\coordinate (Z4) at ({-\yoffset},{0.55 * \scale});
\coordinate (Z5) at ({-\yoffset},{0.95 * \scale});
\coordinate (Z6) at ({-\yoffset},{1.35 * \scale});
\coordinate (Z7) at ({-\yoffset},{1.40 * \scale});

\draw[<->] (Z1) -- (Z2);
\draw[<->] (Z2) -- (Z3);
\draw[<->] (Z3) -- (Z4);
\draw[<->] (Z4) -- (Z5);
\draw[<->] (Z5) -- (Z6);
\draw[<->] (Z6) -- (Z7);

\draw (Z1) -- (Z2) node [midway, above, sloped] (TextNode) {\scriptsize 1};
\draw (Z2) -- (Z3) node [midway, above, sloped] (TextNode) {\scriptsize 0.5};
\draw (Z3) -- (Z4) node [midway, above, sloped] (TextNode) {\scriptsize 4};
\draw (Z4) -- (Z5) node [midway, above, sloped] (TextNode) {\scriptsize 4};
\draw (Z5) -- (Z6) node [midway, above, sloped] (TextNode) {\scriptsize 4};
\draw (Z6) -- (Z7) node [midway, above, sloped] (TextNode) {\scriptsize 0.5};

%\draw (A) -- (B) node [midway, above, sloped] (TextNode) {\footnotesize \SI{300}{\eV}};

\end{tikzpicture}
        \caption{Layout of the Hohlraum problem. All measurements are in \si{\mm}. The filled areas are the dense material, the rest are optically thin.}
        \label{fig:results:hohlraum:design}
    \end{figure}

    The second problem we present in this paper is the heating of a cavity from a radiation source. Similar problems have been studied in literature~\cite{mcclarren_robust_2010,brunner_forms_2002}, but with different geometry and materials. The layout is shown in \cref{fig:results:hohlraum:design}. The problem is \SI{0.65}{\cm} by \SI{1.4}{\cm}, with a square mesh of \(39\times84\) cells. The walls of the cavity have a frequency-dependent opacity
    \begin{equation} \label{eq:results:hohlraum:opacity}
        \opacity\left(\nu, T\right) = \density\alpha\frac{1 - \e{-\frac{\plank\nu}{\boltzmann T}}}{\left(\plank\nu\right)^3},
    \end{equation}
     with the opacity factor \(\alpha = \SI{1e12}{ \electronvolt\cubed\centi\meter\squared\per\gram }\), density \(\density = \SI{1.0}{\gram\per\centi\metre\cubed}\) and the heat capacity \(\cv = \SI{3e12}{erg\per\gram\per\electronvolt }\), while the cavity is filled with a material that is almost a vacuum (\(\density = \SI{1e-3}{\gram\per\centi\metre\cubed}\), \(\cv = \SI{1e12}{ erg\per\gram\per\electronvolt}\)) and highly transparent (\(\opacity = \SI{1e-8}{\per\cm}\)). The frequency was discretized into 100 uniform, logarithmic groups between \(\frequency_\mathrm{min} = \SI{1e-3}{\electronvolt}\) and \(\frequency_\mathrm{max} = \SI{1e6}{\electronvolt}\). The left side of the problem has a reflective boundary condition, all other sides are vacuum conditions. At the beginning, the problem is in thermal equilibrium at \(T_0  = \SI{1.0}{\electronvolt}\), with the temperature at \(y=\SI{0}{\centi\metre}\) set to \(T_\mathrm{inc} = \SI{300}{\electronvolt}\). The problem was run with an initial time step of \(\Delta t_0 = \SI{1e-12}{\second}\), increased by a factor of 1.1 each step up to a maximum of \(\Delta t_\mathrm{max} = \SI{5e-12}{\second}\), for a total time of \(t_\mathrm{end} = \SI{1e-8}{\second}\). We used a tolerance of \(\tau_\mathrm{HOLO} = \SI{1e-4}{}\) for the HOLO solver, \(\tau_\mathrm{P1} = \SI{1e-8}{}\) for the P1 solver and \(\tau_T = \SI{1e-12}{}\) for the temperature solver.
     
    \begin{figure} 
        \begin{minipage}{0.45\textwidth}
            \includegraphics[width=\textwidth]{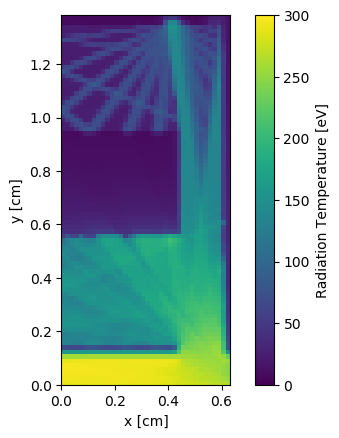}
            \subcaption{Gauss}
            \label{fig:results:hohlraum:temp_rad:gauss}
        \end{minipage} \hfill
        \begin{minipage}{0.45\textwidth}
            \includegraphics[width=\textwidth]{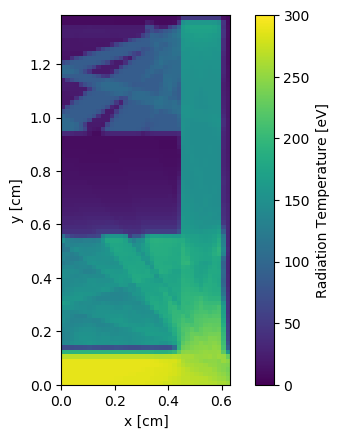}
            \subcaption{Axis aligned}
            \label{fig:results:hohlraum:temp_rad:aligned}
        \end{minipage}
        \begin{minipage}{0.45\textwidth}
            \includegraphics[width=\textwidth]{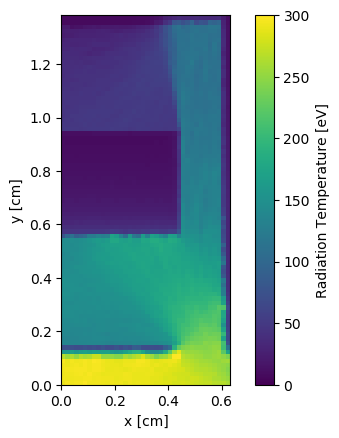}
            \subcaption{Random}
            \label{fig:results:hohlraum:temp_rad:random}
        \end{minipage} \hfill
        \begin{minipage}{0.45\textwidth}
            \includegraphics[width=\textwidth]{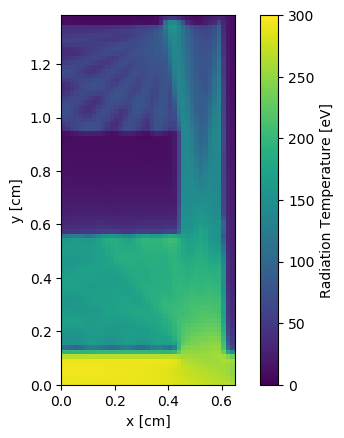}
            \subcaption{Capsaicin}
            \label{fig:results:hohlraum:temp_rad:capsaicin}
        \end{minipage}
        \caption{Radiation temperature for the Hohlraum problem at \SI{1e-8}{s} using the different quadrature types.}
        \label{fig:results:hohlraum:temp_rad}
    \end{figure}

    \begin{figure}
        \begin{minipage}{0.45\textwidth}
            \includegraphics[width=\textwidth]{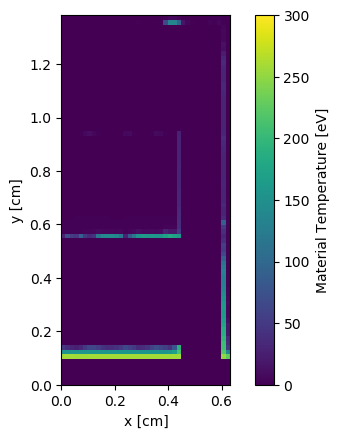}
            \subcaption{Gauss}
            \label{fig:results:hohlraum:temp_mat:gauss}
        \end{minipage} \hfill
        \begin{minipage}{0.45\textwidth}
            \includegraphics[width=\textwidth]{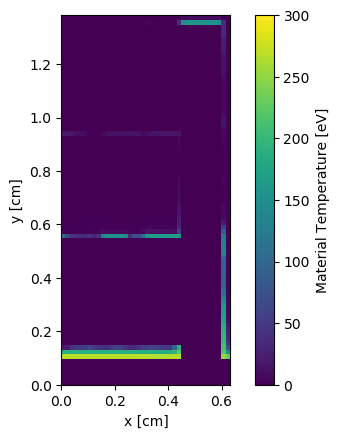}
            \subcaption{Axis Aligned}
            \label{fig:results:hohlraum:temp_mat:aligned}
        \end{minipage}
        \begin{minipage}{0.45\textwidth}
            \includegraphics[width=\textwidth]{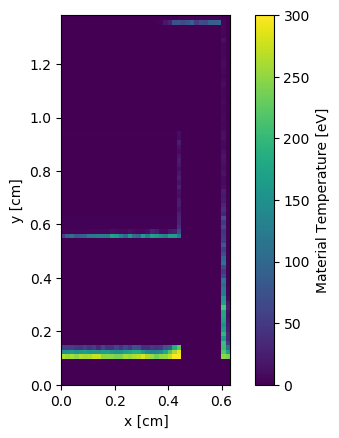}
            \subcaption{Random}
            \label{fig:results:hohlraum:temp_mat:ranomd}
        \end{minipage} \hfill
        \begin{minipage}{0.45\textwidth}
            \includegraphics[width=\textwidth]{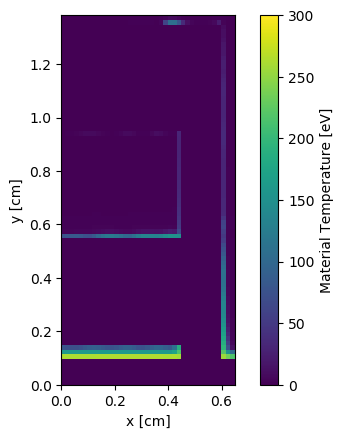}
            \subcaption{Capsaicin}
            \label{fig:results:hohlraum:temp_mat:capsaicin}
        \end{minipage}
        \caption{Material temperature for the Hohlraum problem at \SI{1e-8}{s} using the different quadrature types.}
        \label{fig:results:hohlraum:temp_mat}
    \end{figure}

    The radiation temperature at \(t = \SI{1e-8}{\second}\) is shown in \cref{fig:results:hohlraum:temp_rad:gauss} using a standard \sn Gauss-Chebychev product quadrature with 8 polar and 24 azimuthal angles. This quadrature results a total number of \num{1257984} particles used in the calculation, where we initialize 4
     particles per cell. Note that the number of particles remains constant once initialized, as no particles are killed or created during the calculation. The plot clearly shows ray effects, with heating preferentially occurring along directions included in the quadrature set, while the temperature between these directions stays unphysically cold. They can be seen especially well in the upper region, where the wall sees only a highly localized deposition of energy (as shown by the material temperature in \cref{fig:results:hohlraum:temp_mat:gauss}), whereas it should be a much wider deposition on the part that is not shadowed. The lower wall shows a smooth material temperature profile, while the lower side of the center block already shows indications of ray effects with localized temperature extrema. 
    
    It is customary not to include the axis of the coordinate system in \sn quadrature sets. However, having a quadrature set that is axis-aligned improves the results at the top wall as shown in \cref{fig:results:hohlraum:temp_rad:aligned}. The axis aligned quadrature has 9 polar angles and 20 azimuthal angles, which results in a total of \num{1310400} particles. We chose these settings to be closer to the total number of angles of the standard Gauss-Chebychev quadrature, and to avoid \SI{45}{\degree} angles, which cause many particles to hit the corners of the square-mesh cells. While our implementation is capable of handling particles hitting cell corners, it fails for large numbers of particles because the corner case is highly ill-conditioned due to very short particle tracks. Even though the axis-aligned quadrature improves the results for the top wall, we still see strong ray effects.
    
    To further ameliorate this problem, we switched to a random quadrature with a total number of 96 angles (which is the same as for the Gauss-Chebychev quadrature, since we are two-dimensional). These random angles are different for each particle starting point (4 per cell). After the particles are initialized, they maintain their direction, except when they are reflected at a boundary by the reflection law \cref{eq:method:ho:reflection}. This approach results in a lot more angles covered by the particles, but  introduces random noise. The radiation temperature does not show ray-effects, and a smooth radiation field develops in the optically thin material, as can be seen in \cref{fig:results:hohlraum:temp_rad:random}. However, we see strong differences in the thick material between neighboring cells caused by noise. The material temperature in \cref{fig:results:hohlraum:temp_mat:ranomd} shows this well for the lower wall. Note that the HOLO solver did not converge for the random quadrature case due to abrupt changes in particle surface fluxes when particles cross cells, stalling after about two iterations at a residual magnitude of approximately \num{1e-2}. Therefore we limited the number of HOLO iterations to 5 per time step for the random cases. A test with an increased number of particles per cell showed  a significantly improved HOLO convergence. Future work will explore higher-order particle interpolation to ameliorate this problem.
    
    The results obtained with Capsaicin~\cite{thompson_capsaicin:_2006} are shown in \cref{fig:results:hohlraum:temp_rad:capsaicin,fig:results:hohlraum:temp_mat:capsaicin}. We see good agreement for the radiation temperature in the optically thin material between Capsaicin and the standard Gauss-Chebychev quadrature, \cref{fig:results:hohlraum:temp_rad:gauss}. The ray effects are more smeared out in Capsaicin but clearly visible. The material temperature was higher and more evenly distributed in the upper part of the bottom wall. We believe these differences are caused by Capsaicin's linearization of the $T^4$ nonlinearity in the emission source.
    
    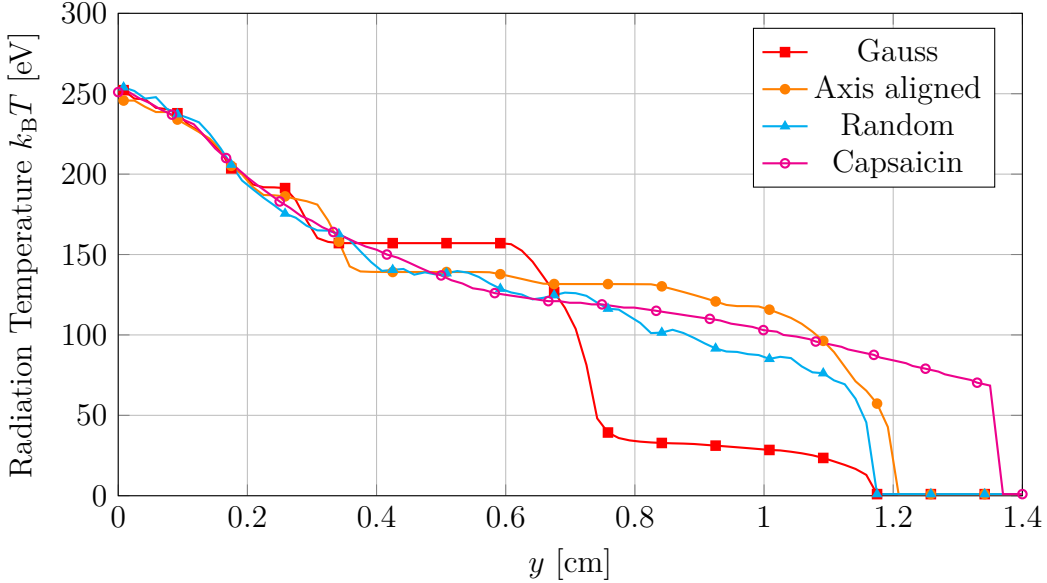
\begin{figure}
        \setlength{\figheight}{8cm}
        \setlength{\figwidth}{0.9\textwidth}
        \begin{tikzpicture}
    \begin{axis}[
            xmin = 0, xmax = 1.4,
            ymin = 0, ymax = 300,
            y tick label style={/pgf/number format/.cd, scaled y ticks = false, fixed},
            x tick label style={/pgf/number format/.cd, scaled x ticks = false, fixed},
            height=\figheight,
            width=\figwidth,
            grid=major,
            legend pos = north east,
            legend columns = 1,
            xlabel = {\(y\) [\si{cm}]},
            ylabel = {Radiation Temperature \(\boltzmann T\) [\si{\electronvolt}]},
            mark options={solid, scale = 0.8, mark repeat = 5},
        ]

        \addplot [red, thick, mark=square*] table [y = y, y = T_rad_10, col sep=comma] {data/hohlraum/hohlraum_gauss_wave.csv};
        \addlegendentry{Gauss}
        \addplot [orange, thick, mark=*] table [y = y, y = T_rad_10, col sep=comma] {data/hohlraum/hohlraum_aligned_wave.csv};
        \addlegendentry{Axis aligned}
        \addplot [cyan, thick, mark=triangle*] table [y = y, y = T_rad_10, col sep=comma] {data/hohlraum/hohlraum_random_wave.csv};
        \addlegendentry{Random}
%        \addplot [teal, thick, mark=diamond*] table [y = y, y = T_rad_10, col sep=comma] {data/hohlraum/hohlraum_gauss_random_wave.csv};
%        \addlegendentry{Gauss Random}
        \addplot [magenta, thick, mark=o] table [y = y, y = T_rad_10, col sep=comma] {data/hohlraum/capsaicin_wave.csv};
        \addlegendentry{Capsaicin}
        
    \end{axis}
\end{tikzpicture}
        \caption{Radiation wave front at \(x = \SI{0.525}{\centi\metre}\) at time \(t = \SI{4e-11}{\second}\) using a constant time step size \(\dt = \SI{1e-11}{\second}\).}
        \label{fig:results:hohlraum:wave}
    \end{figure}

    \Cref{fig:results:hohlraum:wave} shows the radiation wave front at the time \(t = \SI{4e-11}{\second}\) along the dashed line in \cref{fig:results:hohlraum:design}. The exact location should be \(c t = \SI{1.1991}{\centi\metre}\). The results show that our code is within reasonable range of this analytical value. Deviations from this value come from the different angles contained in the quadrature used, i.e., the Gauss quadrature has no direction going perpendicular to the wave front, which results in a slower propagation. The steps in the temperature profile for both the Gauss quadrature and the axis-aligned quadrature are related to the discrete propagation angles considered. The Capsaicin wave front has propagated much further compared to both our results and the analytical value, a consequence of the backward Euler time discretization introducing excessive numerical diffusion.

\FloatBarrier
\subsection{Runtime and Convergence}
    
    To control the runtime necessary for the convergence studies, we limited our study to the one-dimensional Marshak-wave problem. The implementation is the same as in two dimensions.
    
    The Marshak-wave problem propagates a radiation wave through a material with a temperature dependent opacity
    \begin{equation}
        \opacity\left(T\right) = \frac{\density \alpha}{T^3}
    \end{equation}
    with the opacity factor \(\alpha = \SI{1e6}{\electronvolt\cubed\centi\meter\squared\per\gram}\). The problem is \SI{2}{\cm} long divided into \(n\) mesh cells, with vacuum boundaries on both sides. In the beginning, the problem is in thermal equilibrium at \(T_0 = \SI{0.025}{\electronvolt}\), and on the left side a temperature of \(T_\mathrm{inc} = \SI{150}{\electronvolt}\) is applied. The problem was run with an initial time step of \(\Delta t_0 = \SI{1e-12}{\second}\), which increased by a factor of 1.1 each step up to a prescribed maximum \(\Delta t_\mathrm{max}\), for a total time of \(t_\mathrm{end} = \SI{5e-8}{\second}\).
    
    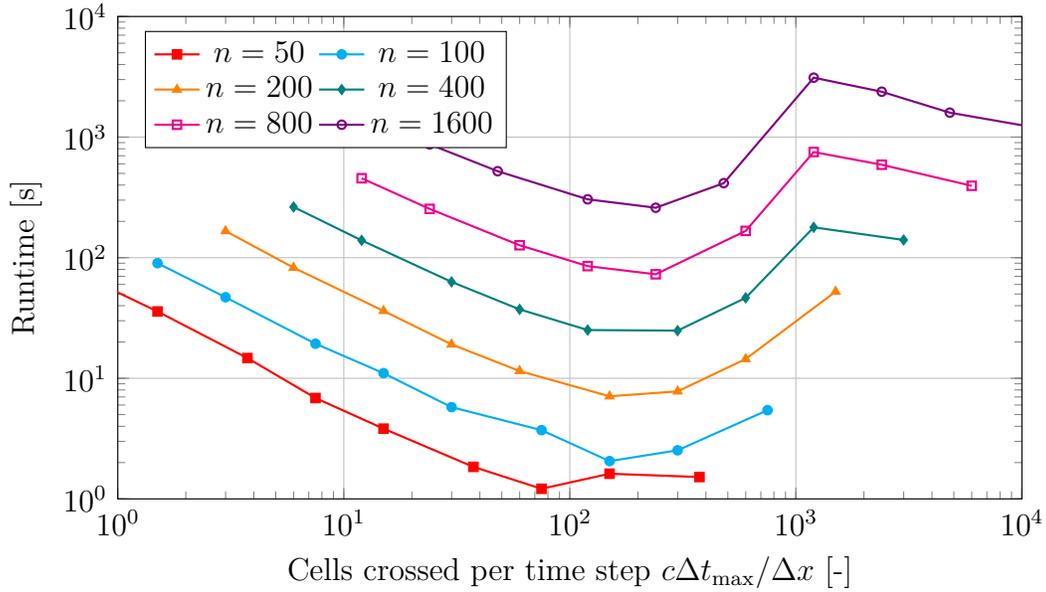
\begin{figure}
        \setlength{\figheight}{8cm}
        \setlength{\figwidth}{0.9\textwidth}
        \begin{tikzpicture}
    \begin{loglogaxis}[
            xmin = 1, xmax = 1e4,
            ymin = 1, ymax = 1e4,
            y tick label style={/pgf/number format/.cd, scaled y ticks = false, fixed},
            x tick label style={/pgf/number format/.cd, scaled x ticks = false, fixed},
            height=\figheight,
            width=\figwidth,
            grid=major,
            legend pos = north west,
            legend columns = 2,
            xlabel = {Cells crossed per time step \({c \Delta t_\mathrm{max}}/{\Delta x}\) [-]},
            ylabel = {Runtime [\si{\second}]},
            mark options={solid, scale = 0.8},
        ]
        
%        \addplot [red, thick, mark=square*] table [x = nx_25_x, y = nx_25_y, col sep=comma] {data/studies/cdt_dx.csv};
%        \addlegendentry{\(n = 25\)}
        
        \addplot [red, thick, mark=square*] table [x = nx_50_x, y = nx_50_y, col sep=comma] {data/studies/cdt_dx.csv};
        \addlegendentry{\(n = 50\)}

        \addplot [cyan, thick, mark=*] table [x = nx_100_x, y = nx_100_y, col sep=comma] {data/studies/cdt_dx.csv};
        \addlegendentry{\(n = 100\)}
        
        \addplot [orange, thick, mark=triangle*] table [x = nx_200_x, y = nx_200_y, col sep=comma] {data/studies/cdt_dx.csv};
        \addlegendentry{\(n = 200\)}
        
        \addplot [teal, thick, mark=diamond*] table [x = nx_400_x, y = nx_400_y, col sep=comma] {data/studies/cdt_dx.csv};
        \addlegendentry{\(n = 400\)}
        
        \addplot [magenta, thick, mark=square] table [x = nx_800_x, y = nx_800_y, col sep=comma] {data/studies/cdt_dx.csv};
        \addlegendentry{\(n = 800\)}
        
%        \addplot [red, thick, mark=square*] table [x = nx_1200_x, y = nx_1200_y, col sep=comma] {data/studies/cdt_dx.csv};
%        \addlegendentry{\(n = 1200\)}
        
        \addplot [violet, thick, mark=o] table [x = nx_1600_x, y = nx_1600_y, col sep=comma] {data/studies/cdt_dx.csv};
        \addlegendentry{\(n = 1600\)}
        
    \end{loglogaxis}
\end{tikzpicture}
        \caption{Runtime as a function of maximal number of cells crossed per time step \( c\Delta t_\mathrm{max} / \Delta x\) for different mesh sizes \(n\).}
        \label{fig:results:convergence:dt_dx}
    \end{figure}

    The total runtime is a function of the maximum time step size \(\Delta t_\mathrm{max}\), as shown in \cref{fig:results:convergence:dt_dx} for different mesh sizes. There are three major effects influencing the runtime. The first factor is the number of time steps required to reach the final time, which decreases inversely with increasing \(\Delta t_\mathrm{max}\). The second is the number of cells a particle crosses on average during one time step, which increases proportionally with the time step size. The last is the number of HOLO iterations necessary to converge the residual. This number increases for large time step sizes, while it remains almost constant for small ones. The combination of these leads to an optimal time step size for which the total run time is minimal. \Cref{fig:results:convergence:dt_dx} shows that this optimum can be found for different mesh sizes for an almost constant ratio of \(c\Delta t_\mathrm{max} / \Delta x\), corresponding to a maximum number of cells crossings between \num{100} and \num{300}. The actual number of cells crossings is a function of the particle's direction.

    \begin{figure}
        \setlength{\figheight}{8cm}
        \setlength{\figwidth}{0.9\textwidth}
        \begin{tikzpicture}
    \begin{axis}[
            xmin = 0, xmax = 2,
            ymin = 0, ymax = 150,
            y tick label style={/pgf/number format/.cd, scaled y ticks = false, fixed},
            x tick label style={/pgf/number format/.cd, scaled x ticks = false, fixed},
            height=\figheight,
            width=\figwidth,
            grid=major,
            legend pos = north east,
            legend columns = 1,
            xlabel = {\(x\) [\si{\cm}] (1600 cells)},
            ylabel = {Material Temperature \(T\) [\si{\electronvolt}]},
            mark options={solid, scale = 0.8, mark repeat = 20},
        ]

        \addplot [red, thick, mark=square*] table [x = x, y = T_mat_1e-12, col sep=comma] {data/studies/cdt_dx_results.csv};
        \addlegendentry{\(\Delta t_\mathrm{max} = \SI{1e-12}{\second}\)}
        
        \addplot [cyan, thick, mark=*] table [x = x, y = T_mat_5e-12, col sep=comma] {data/studies/cdt_dx_results.csv};
        \addlegendentry{\(\Delta t_\mathrm{max} = \SI{5e-12}{\second}\)}
        
        \addplot [orange, thick, mark=triangle*] table [x = x, y = T_mat_1e-11, col sep=comma] {data/studies/cdt_dx_results.csv};
        \addlegendentry{\(\Delta t_\mathrm{max} = \SI{1e-11}{\second}\)}
        
        \addplot [teal, thick, mark=diamond*] table [x = x, y = T_mat_2e-11, col sep=comma] {data/studies/cdt_dx_results.csv};
        \addlegendentry{\(\Delta t_\mathrm{max} = \SI{2e-11}{\second}\)}
        
        \addplot [magenta, thick, mark=square] table [x = x, y = T_mat_5e-11, col sep=comma] {data/studies/cdt_dx_results.csv};
        \addlegendentry{\(\Delta t_\mathrm{max} = \SI{5e-11}{\second}\)}
        
        \addplot [violet, thick, mark=o] table [x = x, y = T_mat_1e-10, col sep=comma] {data/studies/cdt_dx_results.csv};
        \addlegendentry{\(\Delta t_\mathrm{max} = \SI{1e-10}{\second}\)}
        
        \addplot [purple, thick, mark=triangle] table [x = x, y = T_mat_2e-10, col sep=comma] {data/studies/cdt_dx_results.csv};
        \addlegendentry{\(\Delta t_\mathrm{max} = \SI{2e-10}{\second}\)}
        
    \end{axis}
\end{tikzpicture}
        \caption{Results for the thin Marshak wave at \(t = \SI{5e-8}{\second}\) using different \(\Delta t_\mathrm{max}\) and a mesh with 1600 cells.}
        \label{fig:results:convergence:dt_dx_result}
    \end{figure}
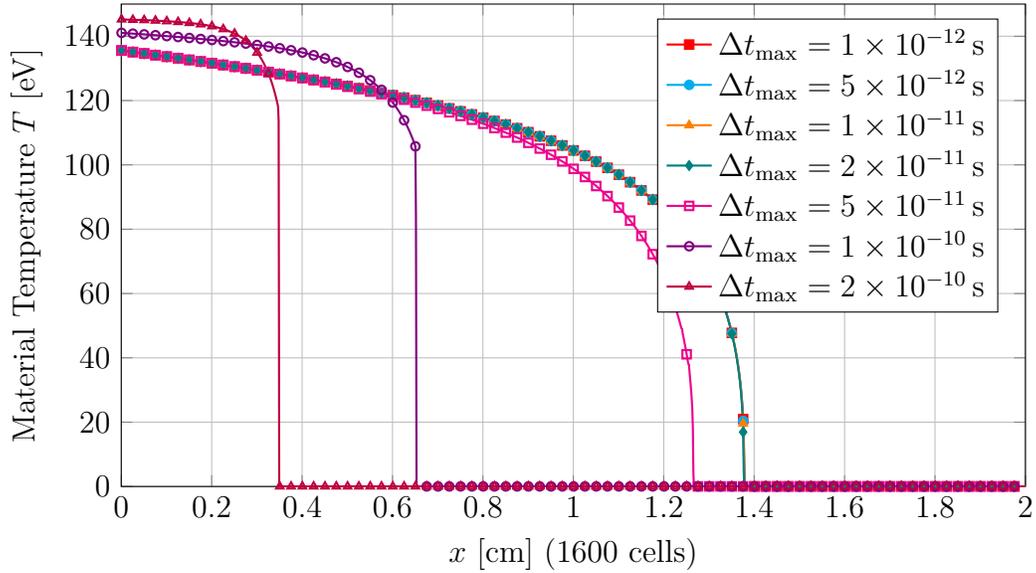

    While the method is stable for large time steps \cite{park_multigroup_2019}, there are upper and lower limits to consider. If the time step is too small, no particle crosses the cell surfaces. This results in zero fluxes for the LO solver, and a decoupling of the cells. This can also occur in a later time step due to alignment of the particles resulting in no surface crossings. The upper time step limit is imposed by the dynamical time scale. If the time step size is too large, the wave front stalls as shown in \cref{fig:results:convergence:dt_dx_result}. For \(\Delta t_\mathrm{max} \le \SI{2e-11}{\second}\), or \(c\Delta t_\mathrm{max} / \Delta x \lesssim 480 \), the wave front reaches the correct final position, but for larger time steps the final position of the wave front lags behind. We observed the same value of the time step threshold for different meshes. 

    \begin{figure}
        \setlength{\figheight}{8cm}
        \setlength{\figwidth}{0.9\textwidth}
        \begin{tikzpicture}
    \begin{axis}[
            xmin = 0, xmax = 140,
            ymin = 0, ymax = 20,
            y tick label style={/pgf/number format/.cd, scaled y ticks = false, fixed},
            x tick label style={/pgf/number format/.cd, scaled x ticks = false, fixed},
            height=\figheight,
            width=\figwidth,
            grid=major,
            legend pos = north west,
            legend columns = 1,
            xlabel = {Number of groups [-]},
            ylabel = {Relative runtime [-]},
            mark options={solid, scale = 0.8},
        ]

        \addplot [red, thick, mark=square*] table [x = groups, y = dt_time_100, col sep=comma] {data/studies/groups.csv};
        \addlegendentry{\(n = 100\)}
        
        \addplot [cyan, thick, mark=*] table [x = groups, y = dt_time_200, col sep=comma] {data/studies/groups.csv};
        \addlegendentry{\(n = 200\)}
        
        \addplot [orange, thick, mark=triangle*] table [x = groups, y = dt_time_400, col sep=comma] {data/studies/groups.csv};
        \addlegendentry{\(n = 400\)}
        
    \end{axis}
\end{tikzpicture}
        \caption{Runtime as a function of number of groups for different numbers of mesh cells \(n\).}
        \label{fig:results:convergence:groups}
    \end{figure}
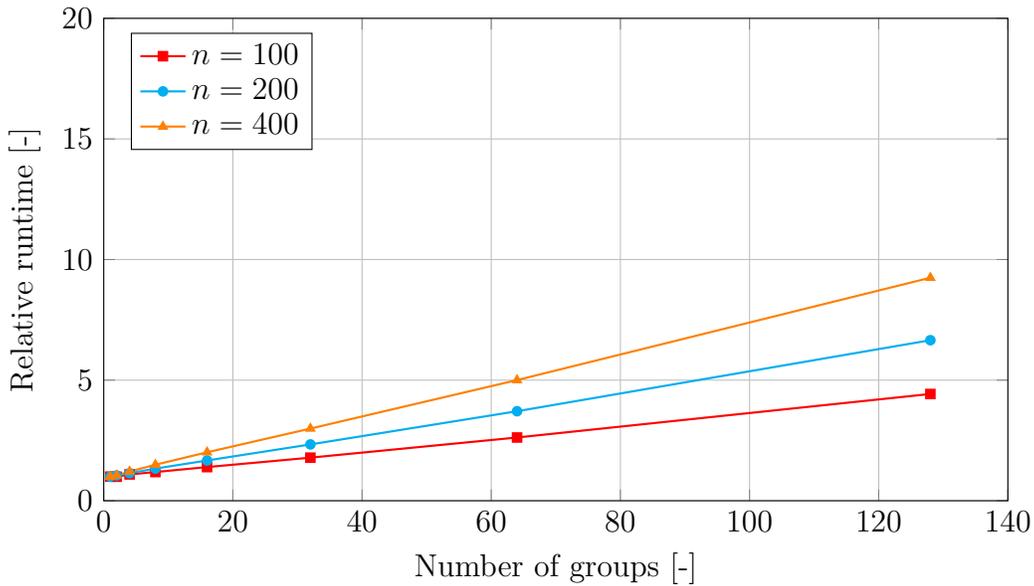

    The DP method is very well suited for multi-frequency calculations as mentioned before. All frequency information is carried by each particle, and the group iteration is the innermost loop during the tracking step. Therefore, the runtime is proportional to the number of frequency groups with a factor significantly less than unity. \Cref{fig:results:convergence:groups} shows the runtime of the Marshak wave problem as a function of group number for several mesh sizes using \(\Delta t_\mathrm{max} = \SI{1e-12}{\second}\). The slope is dependent on the ratio of work done within the inner loop to the work outside of it. A finer mesh results in more cell crossings, increasing the ratio and hence increasing the slope. However, the slope will always remain less than unity.

    \begin{figure}
        \setlength{\figheight}{8cm}
        \setlength{\figwidth}{0.9\textwidth}
        \begin{tikzpicture}
    \begin{loglogaxis}[
            xmin = 10, xmax = 2e3,
            ymin = 1e-3, ymax = 0.2,
            y tick label style={/pgf/number format/.cd, scaled y ticks = false, fixed},
            x tick label style={/pgf/number format/.cd, scaled x ticks = false, fixed},
            height=\figheight,
            width=\figwidth,
            grid=major,
            legend pos = north east,
            legend columns = 2,
            xlabel = {Number of cells [-]},
            ylabel = {Relative error [-]},
            mark options={solid, scale = 0.8},
        ]
        
%        \addplot [red, thick, mark=square*] table [x = nx_25_x, y = nx_25_y, col sep=comma] {data/studies/cdt_dx.csv};
%        \addlegendentry{\(n = 25\)}
        
        \addplot [red, thick, mark=square*] table [x = n, y = 2, col sep=comma] {data/studies/spatial.csv};
        \addlegendentry{\(M_\mathrm{P} = 2\)}

        \addplot [cyan, thick, mark=*] table [x = n, y = 4, col sep=comma] {data/studies/spatial.csv};
        \addlegendentry{\(M_\mathrm{P} = 4\)}
        \addplot [orange, thick, mark=triangle*] table [x = n, y = 8, col sep=comma] {data/studies/spatial.csv};
        \addlegendentry{\(M_\mathrm{P} = 8\)}
        
        \addplot [teal, thick, mark=diamond*] table [x = n, y = 16, col sep=comma] {data/studies/spatial.csv};
        \addlegendentry{\(M_\mathrm{P} = 16\)}
        
        \addplot [magenta, thick, mark=square] table [x = n, y = 32, col sep=comma] {data/studies/spatial.csv};
        \addlegendentry{\(M_\mathrm{P} = 32\)}
        
        \addplot [black, thick, dashed] table [x = n, y expr= \thisrow{n} ^ -1, col sep=comma] {data/studies/spatial.csv};
        \addlegendentry{\(O\left(x\right)\)}
        
    \end{loglogaxis}
\end{tikzpicture}
        \caption{Error convergence with the number of mesh cells for different number of polar angles \(M_\mathrm{P}\).}
        \label{fig:results:convergence:spatial}
    \end{figure}
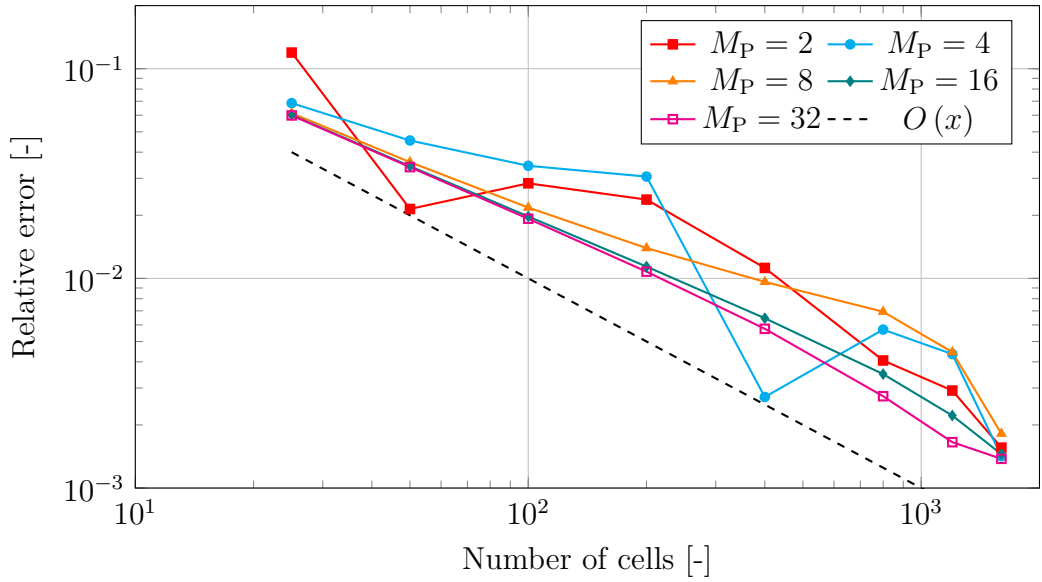

    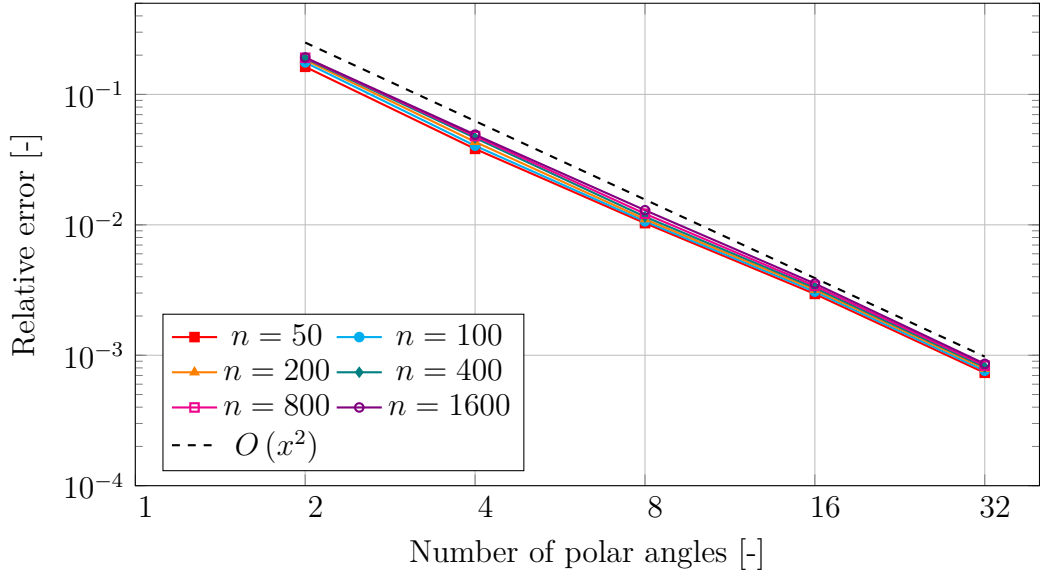
\begin{figure}
        \setlength{\figheight}{8cm}
        \setlength{\figwidth}{0.9\textwidth}
        \begin{tikzpicture}
    \begin{loglogaxis}[
            xmin = 1, xmax = 40,
            ymin = 1e-4, ymax = 0.5,
            y tick label style={/pgf/number format/.cd, scaled y ticks = false, fixed},
            x tick label style={/pgf/number format/.cd, scaled x ticks = false, fixed},
            log x ticks with fixed point,
            xtick = {1, 2, 4, 8, 16, 32},
            height=\figheight,
            width=\figwidth,
            grid=major,
            legend pos = south west,
            legend columns = 2,
            xlabel = {Number of polar angles [-]},
            ylabel = {Relative error [-]},
            mark options={solid, scale = 0.8},
        ]
        
%        \addplot [red, thick, mark=square*] table [x = nx_25_x, y = nx_25_y, col sep=comma] {data/studies/cdt_dx.csv};
%        \addlegendentry{\(n = 25\)}
        
        \addplot [red, thick, mark=square*] table [x = angles, y = 50, col sep=comma] {data/studies/polar.csv};
        \addlegendentry{\(n =  50\)}

        \addplot [cyan, thick, mark=*] table [x = angles, y = 100, col sep=comma] {data/studies/polar.csv};
        \addlegendentry{\(n =  100\)}
        \addplot [orange, thick, mark=triangle*] table [x = angles, y = 200, col sep=comma] {data/studies/polar.csv};
        \addlegendentry{\(n =  200\)}
        
        \addplot [teal, thick, mark=diamond*] table [x = angles, y = 400, col sep=comma] {data/studies/polar.csv};
        \addlegendentry{\(n =  400\)}
        
        \addplot [magenta, thick, mark=square]table [x = angles, y = 800, col sep=comma] {data/studies/polar.csv};
        \addlegendentry{\(n =  800\)}
        
        \addplot [violet, thick, mark=o]table [x = angles, y = 1600, col sep=comma] {data/studies/polar.csv};
        \addlegendentry{\(n =  1600\)}
        
        \addplot [black, thick, dashed] table [x = angles, y expr= \thisrow{angles} ^ -2, col sep=comma] {data/studies/polar.csv};
        \addlegendentry{\(O\left(x^2\right)\)}
        
    \end{loglogaxis}
\end{tikzpicture}
        \caption{Error convergence with the number of polar angles for different number of mesh cells \(n\).}
        \label{fig:results:convergence:polar}
    \end{figure}

    \begin{figure}
        \setlength{\figheight}{8cm}
        \setlength{\figwidth}{0.9\textwidth}
        \begin{tikzpicture}
    \begin{loglogaxis}[
            xmin = 1, xmax = 20,
            ymin = 1e-8, ymax = 1e-4,
            y tick label style={/pgf/number format/.cd, scaled y ticks = false, fixed},
            x tick label style={/pgf/number format/.cd, scaled x ticks = false, fixed},
            log x ticks with fixed point,
            xtick = {1, 2, 4, 8, 16, 32},
            height=\figheight,
            width=\figwidth,
            grid=major,
            legend pos = south west,
            legend columns = 3,
            xlabel = {Number of particles per cell and direction [-]},
            ylabel = {Relative error [-]},
            mark options={solid, scale = 0.8},
        ]
        
%        \addplot [red, thick, mark=square*] table [x = nx_25_x, y = nx_25_y, col sep=comma] {data/studies/cdt_dx.csv};
%        \addlegendentry{\(n = 25\)}
        
        \addplot [red, thick, mark=square*] table [x = particles, y = 50, col sep=comma] {data/studies/particles.csv};
        \addlegendentry{\(n =  50\)}

        \addplot [cyan, thick, mark=*] table [x = particles, y = 100, col sep=comma] {data/studies/particles.csv};
        \addlegendentry{\(n =  100\)}
        \addplot [orange, thick, mark=triangle*] table [x = particles, y = 200, col sep=comma] {data/studies/particles.csv};
        \addlegendentry{\(n =  200\)}
        
        \addplot [teal, thick, mark=diamond*] table [x = particles, y = 400, col sep=comma] {data/studies/particles.csv};
        \addlegendentry{\(n =  400\)}
        
        \addplot [magenta, thick, mark=square]table [x = particles, y = 800, col sep=comma] {data/studies/particles.csv};
        \addlegendentry{\(n =  800\)}
        
        \addplot [violet, thick, mark=o]table [x = particles, y = 1600, col sep=comma] {data/studies/particles.csv};
        \addlegendentry{\(n =  1600\)}
        
        \addplot [black, thick, dashed] table [x = particles, y expr=5e-5 *  \thisrow{particles} ^ -1, col sep=comma] {data/studies/particles.csv};
        \addlegendentry{\(O\left(x\right)\)}
        
    \end{loglogaxis}
\end{tikzpicture}
        \caption{Error convergence with the number of  particle starting points per cell with 32 polar angles per point and different number of mesh cells \(n\).}
        \label{fig:results:convergence:particles}
    \end{figure}
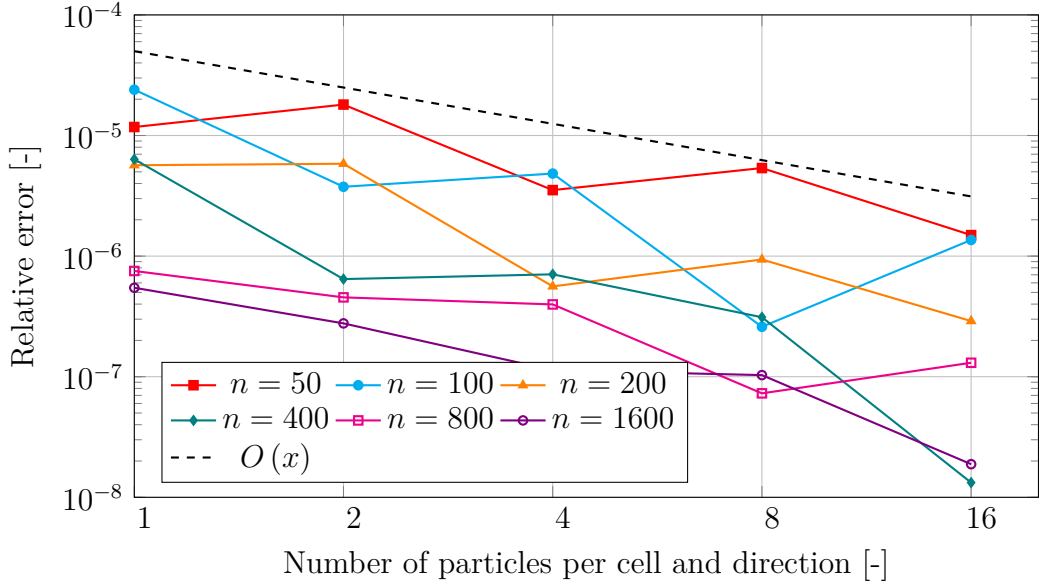

    We present next results on the convergence properties of the algorithm. To reduce the runtime for the convergence studies, we limited the simulation time to \(t_\mathrm{end} = \SI{2e-9}{\second}\), and the problem size to \SI{1}{\cm}. To avoid problems with the upper time step-size limit, we used a time step of \(\Delta t =\SI{1e-12}{\second}\).  For the spatial convergence, we use a mesh with \(n = 3200\) cells  as reference to calculate the error. The results shown in \cref{fig:results:convergence:spatial} indicate an approximately first-order convergence rate with spatial refinement. The convergence in angle is second order, as can be seen in \cref{fig:results:convergence:polar}, where the reference solution used \(M_\mathrm{P} = 64\) polar angles. \Cref{fig:results:convergence:particles} shows the convergence with the number of particle starting points per cell (each point uses all directions of the angular quadrature). The error with respect to a reference solution using 32 starting points is much lower than for the other cases, \cref{fig:results:convergence:spatial,fig:results:convergence:polar}, and converges with first order. However, there seem to be some fluctuations.

\FloatBarrier
\section{Conclusion}

    We have extended the moment accelerated, multi-frequency deterministic particle method proposed by \citet{park_multigroup_2019} to two dimensional thermal radiative transfer problems using a fast ray-tracing algorithm. A linear reconstruction of the emission source, obtained from a discretely consistent, moment-based low-order solver, allows the analytical integration of the characteristic equation along a particle track, resulting in an improved solution in optically thick materials compared to other particle methods. In contrast to Monte-Carlo (MC) methods, our method does not feature randomness (with the possible exception of particle initialization and random quadrature to ameliorate ray effects), therefore the solution does not contain stochastic noise. We showed how we can obtain a linear reconstruction of the energy deposition, material temperature and emission source using spatial moments. With this, we were able to demonstrate that our HO system features the asymptotic diffusion limit, at least on triangular meshes.
    
    We further showed that, using the flexibility of a particle method, we can reduce ray-effects by using random quadrature sets, but it introduces noise in the solution.
    
    Future work will include the implementation of a bi-linear source reconstruction for rectangular meshes, the development of a DG discretization scheme for the LO system, which self-consistently solves for the LO slope, and the extension to cylindrical geometries.
    
\section{Acknowledgment}

    This work was supported by the US Department of Energy through the Los Alamos National Laboratory. Los Alamos National Laboratory is operated by Triad National Security, LLC, for the National Nuclear Security Administration of U.S. Department of Energy (Contract No. 89233218CNA000001). Research presented in this article was supported by the Laboratory Directed Research and Development program of Los Alamos National Laboratory under project number 20160448ER.
    
    \FloatBarrier
    
\section{References}
    \bibliographystyle{plainnat}
    \bibliography{literature}

\begin{thebibliography}{26}
\providecommand{\natexlab}[1]{#1}
\providecommand{\url}[1]{\texttt{#1}}
\expandafter\ifx\csname urlstyle\endcsname\relax
  \providecommand{\doi}[1]{doi: #1}\else
  \providecommand{\doi}{doi: \begingroup \urlstyle{rm}\Url}\fi

\bibitem[Adams(2001)]{adams_discontinuous_2001}
Marvin~L. Adams.
\newblock Discontinuous {{Finite Element Transport Solutions}} in {{Thick
  Diffusive Problems}}.
\newblock \emph{Nuclear Science and Engineering}, 137\penalty0 (3):\penalty0
  298--333, March 2001.
\newblock ISSN 0029-5639.
\newblock \doi{10.13182/NSE00-41}.
\newblock URL \url{https://doi.org/10.13182/NSE00-41}.

\bibitem[Adams et~al.(1998)Adams, Wareing, and
  Walters]{adams_characteristic_1998}
Marvin~L. Adams, Todd~A. Wareing, and Wallace~F. Walters.
\newblock Characteristic {{Methods}} in {{Thick Diffusive Problems}}.
\newblock \emph{Nuclear Science and Engineering}, 130\penalty0 (1):\penalty0
  18--46, September 1998.
\newblock ISSN 0029-5639.
\newblock \doi{10.13182/NSE98-A1987}.
\newblock URL \url{https://doi.org/10.13182/NSE98-A1987}.

\bibitem[Brunner(2002)]{brunner_forms_2002}
Thomas~A Brunner.
\newblock Forms of {{Approximate Radiation Transport}}.
\newblock Technical Report SAND2002-1778, 800993, {Sandia National
  Laboratories}, Albuquerque, NM (United States), June 2002.
\newblock URL \url{http://www.osti.gov/servlets/purl/800993/}.

\bibitem[Chac\'on et~al.(2017)Chac\'on, Chen, Knoll, Newman, Park, Taitano,
  Willert, and Womeldorff]{chacon_multiscale_2017}
L.~Chac\'on, G.~Chen, D.~A. Knoll, C.~Newman, H.~Park, W.~Taitano, J.~A.
  Willert, and G.~Womeldorff.
\newblock Multiscale high-order/low-order ({{HOLO}}) algorithms and
  applications.
\newblock \emph{Journal of Computational Physics}, 330:\penalty0 21--45,
  February 2017.
\newblock ISSN 0021-9991.
\newblock \doi{10.1016/j.jcp.2016.10.069}.
\newblock URL
  \url{http://www.sciencedirect.com/science/article/pii/S0021999116305770}.

\bibitem[Ferrer et~al.(2012)Ferrer, Rhodes, and Smith]{ferrer_linear_2012}
R.~Ferrer, J.~Rhodes, and K.~Smith.
\newblock Linear source approximation in {{CASMO5}}.
\newblock In \emph{{{PHYSOR}} 2012 - {{Advances}} in {{Reactor Physics}} -
  {{Linking Research}}, {{Industry}} and {{Education}}}, Knoxville, TN, April
  2012. {American Nuclear Society}.
\newblock URL \url{http://inis.iaea.org/Search/search.aspx?orig_q=RN:44063435}.

\bibitem[Ferrer and Rhodes~III(2016)]{ferrer_linear_2016}
Rodolfo~M. Ferrer and Joel~D. Rhodes~III.
\newblock A {{Linear Source Approximation Scheme}} for the {{Method}} of
  {{Characteristics}}.
\newblock \emph{Nuclear Science and Engineering}, 182\penalty0 (2):\penalty0
  151--165, February 2016.
\newblock ISSN 0029-5639.
\newblock \doi{10.13182/NSE15-6}.
\newblock URL \url{https://doi.org/10.13182/NSE15-6}.

\bibitem[Ferrer and Rhodes~III(2018)]{ferrer_linear_2018}
Rodolfo~M. Ferrer and Joel~D. Rhodes~III.
\newblock The linear source approximation and particle conservation in the
  {{Method}} of {{Characteristics}} for isotropic and anisotropic sources.
\newblock \emph{Annals of Nuclear Energy}, 115:\penalty0 209--219, May 2018.
\newblock ISSN 0306-4549.
\newblock \doi{10.1016/j.anucene.2018.01.023}.
\newblock URL
  \url{https://www.sciencedirect.com/science/article/pii/S0306454918300203}.

\bibitem[Fleck and Cummings(1971)]{fleck_implicit_1971}
J.~A. Fleck and J.~D. Cummings.
\newblock An implicit {{Monte Carlo}} scheme for calculating time and frequency
  dependent nonlinear radiation transport.
\newblock \emph{Journal of Computational Physics}, 8\penalty0 (3):\penalty0
  313--342, December 1971.
\newblock ISSN 0021-9991.
\newblock \doi{10.1016/0021-9991(71)90015-5}.
\newblock URL
  \url{http://www.sciencedirect.com/science/article/pii/0021999171900155}.

\bibitem[Gentile(2001)]{gentile_implicit_2001}
N.~A. Gentile.
\newblock Implicit {{Monte Carlo Diffusion}}\textemdash{{An Acceleration
  Method}} for {{Monte Carlo Time}}-{{Dependent Radiative Transfer
  Simulations}}.
\newblock \emph{Journal of Computational Physics}, 172\penalty0 (2):\penalty0
  543--571, September 2001.
\newblock ISSN 0021-9991.
\newblock \doi{10.1006/jcph.2001.6836}.
\newblock URL
  \url{http://www.sciencedirect.com/science/article/pii/S0021999101968366}.

\bibitem[Glassner(1989)]{glassner_introduction_1989}
Andrew~S. Glassner.
\newblock \emph{An {{Introduction}} to {{Ray Tracing}}}.
\newblock {Elsevier}, June 1989.
\newblock ISBN 978-0-08-049905-5.

\bibitem[Gol'din(1964)]{goldin_quasi-diffusion_1964}
V.~Ya. Gol'din.
\newblock A quasi-diffusion method of solving the kinetic equation.
\newblock \emph{USSR Computational Mathematics and Mathematical Physics},
  4\penalty0 (6):\penalty0 136--149, January 1964.
\newblock ISSN 00415553.
\newblock \doi{10.1016/0041-5553(64)90085-0}.
\newblock URL
  \url{http://linkinghub.elsevier.com/retrieve/pii/0041555364900850}.

\bibitem[Hammer et~al.(2018)Hammer, Park, and
  Chac\'on]{hammer_multi-dimensional_2018}
Hans Hammer, HyeongKae Park, and Luis Chac\'on.
\newblock A multi-dimensional, moment-accelerated deterministic particle method
  for time-dependent thermal radiative transfer problems.
\newblock In \emph{{{ANS Annual Winter Meeting}} 2018}, Orlanda, Florida,
  November 2018. {ANS}.

\bibitem[Larsen et~al.(1983)Larsen, Pomraning, and
  Badham]{larsen_asymptotic_1983}
E.~W. Larsen, G.~C. Pomraning, and V.~C. Badham.
\newblock Asymptotic analysis of radiative transfer problems.
\newblock \emph{Journal of Quantitative Spectroscopy and Radiative Transfer},
  29\penalty0 (4):\penalty0 285--310, April 1983.
\newblock ISSN 0022-4073.
\newblock \doi{10.1016/0022-4073(83)90048-1}.
\newblock URL
  \url{http://www.sciencedirect.com/science/article/pii/0022407383900481}.

\bibitem[Larsen and Morel(1989)]{larsen_asymptotic_1989}
Edward~W Larsen and J.~E. Morel.
\newblock Asymptotic solutions of numerical transport problems in optically
  thick, diffusive regimes {{II}}.
\newblock \emph{Journal of Computational Physics}, 83\penalty0 (1):\penalty0
  212--236, July 1989.
\newblock ISSN 0021-9991.
\newblock \doi{10.1016/0021-9991(89)90229-5}.
\newblock URL
  \url{http://www.sciencedirect.com/science/article/pii/0021999189902295}.

\bibitem[Larsen et~al.(1987)Larsen, Morel, and
  Miller~Jr.]{larsen_asymptotic_1987}
Edward~W Larsen, J.~E Morel, and Warren~F Miller~Jr.
\newblock Asymptotic solutions of numerical transport problems in optically
  thick, diffusive regimes.
\newblock \emph{Journal of Computational Physics}, 69\penalty0 (2):\penalty0
  283--324, April 1987.
\newblock ISSN 0021-9991.
\newblock \doi{10.1016/0021-9991(87)90170-7}.
\newblock URL
  \url{http://www.sciencedirect.com/science/article/pii/0021999187901707}.

\bibitem[McClarren and Hauck(2010)]{mcclarren_robust_2010}
Ryan~G. McClarren and Cory~D. Hauck.
\newblock Robust and accurate filtered spherical harmonics expansions for
  radiative transfer.
\newblock \emph{Journal of Computational Physics}, 229\penalty0 (16):\penalty0
  5597--5614, August 2010.
\newblock ISSN 0021-9991.
\newblock \doi{10.1016/j.jcp.2010.03.043}.
\newblock URL
  \url{http://www.sciencedirect.com/science/article/pii/S0021999110001622}.

\bibitem[Pandya and Adams(2009)]{pandya_method_2009}
T.~M. Pandya and M.~L. Adams.
\newblock Method of long characteristics applied in space and time.
\newblock In \emph{International {{Conference}} on {{Mathematics}},
  {{Computational Methods}} \& {{Reactor Physics}} ({{M}}\&{{C}} 2009)},
  Saratoga Springs, NY, May 2009.

\bibitem[Pandya et~al.(2011)Pandya, Adams, and Hawkins]{pandya_long_2011}
Tara~M. Pandya, Marvin~L. Adams, and W.~Daryl Hawkins.
\newblock Long characteristics with piecewise linear sources designed for
  unstructured grids.
\newblock In \emph{International {{Conference}} on {{Mathematics}} and
  {{Computational Methods Applied}} to {{Nuclear Science}} and {{Engineering}}
  ({{M}}\&{{C}} 2011)}, Rio de Janeiro, RJ, Brazil, May 2011. {Latin American
  Section (LAS) / American Nuclear Society (ANS)}.
\newblock ISBN 978-85-63688-00-2.
\newblock URL \url{http://inis.iaea.org/Search/search.aspx?orig_q=RN:48022307}.

\bibitem[Park et~al.(2012)Park, Knoll, Rauenzahn, Wollaber, and
  Densmore]{park_consistent_2012}
H.~Park, D.~A. Knoll, R.~M. Rauenzahn, A.~B. Wollaber, and J.~D. Densmore.
\newblock A {{Consistent}}, {{Moment}}-{{Based}}, {{Multiscale Solution
  Approach}} for {{Thermal Radiative Transfer Problems}}.
\newblock \emph{Transport Theory and Statistical Physics}, 41\penalty0
  (3-4):\penalty0 284--303, May 2012.
\newblock ISSN 0041-1450.
\newblock \doi{10.1080/00411450.2012.671224}.
\newblock URL \url{http://dx.doi.org/10.1080/00411450.2012.671224}.

\bibitem[Park et~al.(2013)Park, Knoll, Rauenzahn, Newman, Densmore, and
  Wollaber]{park_efficient_2013}
H.~Park, D.~Knoll, R.~Rauenzahn, C.~Newman, J.~Densmore, and A.~Wollaber.
\newblock An {{Efficient}} and {{Time Accurate}}, {{Moment}}-{{Based
  Scale}}-{{Bridging Algorithm}} for {{Thermal Radiative Transfer Problems}}.
\newblock \emph{SIAM Journal on Scientific Computing}, 35\penalty0
  (5):\penalty0 S18--S41, January 2013.
\newblock ISSN 1064-8275.
\newblock \doi{10.1137/120881075}.
\newblock URL \url{http://epubs.siam.org/doi/abs/10.1137/120881075}.

\bibitem[Park et~al.(2019)Park, Chac\'on, Matsekh, and
  Chen]{park_multigroup_2019}
H.~Park, L.~Chac\'on, A.~Matsekh, and G.~Chen.
\newblock A {{Multigroup Moment}}-{{Accelerated Deterministic Particle Solver}}
  for {{Time}}-dependent {{Thermal Radiative Transfer Problems}}.
\newblock \emph{arXiv:1901.05454 [physics]}, January 2019.
\newblock URL \url{http://arxiv.org/abs/1901.05454}.

\bibitem[Park et~al.(2018)Park, Chac\'on, Matsekh, Chen, and
  Hammer]{park_multigroup_2018-1}
HyeongKae Park, Luis Chac\'on, Anna Matsekh, Guangye Chen, and Hans Hammer.
\newblock Multigroup {{Deterministic Particle Solver}} for {{1D Curvilinear
  Geometries}}.
\newblock In \emph{{{ANS Annual Winter Meeting}} 2018}, Orlanda, Florida,
  November 2018. {ANS}.

\bibitem[Steger(1996)]{steger_calculation_1996}
Carsten Steger.
\newblock On the calculation of moments of polygons.
\newblock Technical report, {Technical Report
  FGBV\textendash{}96\textendash{}04, Forschungsgruppe Bildverstehen (FG BV),
  Informatik IX, Technische Universit\"at M\"unchen}, 1996.

\bibitem[Thompson et~al.(2006)Thompson, Warsa, Budge, and
  Chang]{thompson_capsaicin:_2006}
Kelly~G. Thompson, James~S. Warsa, Kent~G Budge, and Jae~H Chang.
\newblock Capsaicin: {{Deterministic Thermal Radiative Transfer}}.
\newblock Technical Report LA-UR-06-3387, {Los Alamos National Laboratory}, Los
  Alamos, NM, USA, May 2006.

\bibitem[Wollaeger et~al.(2016)Wollaeger, Wollaber, Urbatsch, and
  Densmore]{wollaeger_implicit_2016}
Ryan~T. Wollaeger, Allan~B. Wollaber, Todd~J. Urbatsch, and Jeffery~D.
  Densmore.
\newblock Implicit {{Monte Carlo}} with a {{Linear Discontinuous Finite Element
  Material Solution}} and {{Piecewise Non}}-{{Constant Opacity}}.
\newblock \emph{Journal of Computational and Theoretical Transport},
  45\penalty0 (1-2):\penalty0 123--157, February 2016.
\newblock ISSN 2332-4309.
\newblock \doi{10.1080/23324309.2016.1157491}.
\newblock URL \url{https://doi.org/10.1080/23324309.2016.1157491}.

\bibitem[Yee et~al.(2017)Yee, Wollaber, Haut, and Park]{yee_stable_2017}
B.~C. Yee, A.~B. Wollaber, T.~S. Haut, and H.~Park.
\newblock A {{Stable 1D Multigroup High}}-{{Order Low}}-{{Order Method}}.
\newblock \emph{Journal of Computational and Theoretical Transport},
  46\penalty0 (1):\penalty0 46--76, July 2017.
\newblock ISSN 2332-4309.
\newblock \doi{10.1080/23324309.2016.1187172}.
\newblock URL \url{https://doi.org/10.1080/23324309.2016.1187172}.

\end{thebibliography}
    
\begin{appendices}
\crefalias{section}{appsec}
\crefalias{subsection}{appsec}

\section{Appendix}

\subsection{Low-Order solution strategy} \label{sec:appendix:lo_solution}

    The LO system is solved using a Newton-Krylov method with nonlinear elimination \cite{park_consistent_2012}. Given an iterate for the radiation energy density \(\avg{E}_{i,n,\ell}\) at LO iteration \(\ell\), the radiative flux can be calculated from
    \begin{equation}
    \avg{F}_{j,n,\ell+1} = \frac{1}{\frac{1}{c\Delta t_n} + \opacity[j,n]} \left(
    \left(\gamma^{+}_{j,n} + \frac{1}{3 \Delta x_{j}} \right) c\avg{E}_{j-\half,n,\ell}  
    - \left(\gamma^{-}_{j,n} + \frac{1}{3\Delta x_{j}}\right) c\avg{E}_{j+\half,n,\ell} 
    + \frac{\avg{F}_{j,n-1}}{c\Delta t_n} \right)
    \end{equation}
    and the material temperature 
    \begin{equation}
    \density \cv \frac{T_{i,n,\ell+1} - T_{i,n-1}}{\Delta t_n} + \opacity[i,n] ac T_{i,n,\ell+1}^4 -  \opacity[i,n] c\avg{E}_{i,n,\ell} = 0.
    \end{equation} 
    using a cell-wise Newton solve.
    With the new iterates, we can update the radiation energy density with a Newton-Krylov step. The residual vector \(\vec{R}_{n, \ell+1}\)
    \begin{subequations}
        \begin{equation}
        R_{i,n,\ell+1} = \frac{\avg{E}_{i,n,\ell} - \avg{E}_{i,n-1}}{\Delta t_n} + \sum_{j \in i}\normal_{ij} \vd \normal_{j}\frac{\avg{F}_{j,n,\ell+1} A_j}{V_{i}} + \opacity[i,n] c \avg{E}_{i,n,\ell} - \opacity[i,n] ac T_{i,n,\ell+1}^{4} - \mathcal{R}_{i,n}
        \end{equation}
        is used to solve for the Newton update
        \begin{equation} \label{eq:mathod:lo:newton_update}
        \delta_{n, \ell+1} = -\tilde{\jacobian}_{n, \ell+1}^{-1} \vec{R}_{n,\ell+1}
        \end{equation}
        and to update the solution vector for the radiation energy density
        \begin{equation}
        \vec{E}_{n,\ell+1} = \vec{E}_{n,\ell} + \delta_{n,\ell+1}.
        \end{equation}
    \end{subequations}
    The Jacobian in \cref{eq:mathod:lo:newton_update} is given by
    \begin{equation}
    \tilde{\jacobian} = \jacobian_{EE} - \jacobian_{ET} \jacobian_{TT}^{-1} \jacobian_{TE} - \jacobian_{EF} \jacobian_{FF}^{-1} \jacobian_{FE}
    \end{equation}
    which is found via Gauss Block elimination of the LO Jacobian
    \begin{equation}
    \jacobian = \left[
    \begin{matrix}
    \jacobian_{EE} & \jacobian_{EF} & \jacobian_{ET} \\
    \jacobian_{FE} & \jacobian_{FF} & 0            \\
    \jacobian_{TE} & 0            & \jacobian_{TT}
    \end{matrix}
    \right].
    \end{equation}
    The submatrices can be split into two groups, diagonal matrices
    \begin{subequations}
        \begin{align}
        \jacobian_{EE,ii} &= \left(\frac{1}{\Delta t_n} + \opacityE[i,n] c \right) \\
        \jacobian_{ET,ii} &= -4\opacityP[i,n] ac T_{i,n,\ell+1}^{3} \\
        \jacobian_{TT,ii} &= \left(\frac{\density_{i} c_{v,i}}{\Delta t_n} + 4\opacityP[i,n]acT_{i,n,\ell+1}^{3} \right) \\
        \jacobian_{TE,ii} &= -\opacityE[i,n] c \\
        \jacobian_{FF,ii} &= \frac{1}{c\Delta t_{n}} + \opacityR[j,n]
        \end{align}
        and matrices with off-diagonal parts
        \begin{align}
        \jacobian_{EF,ij} &= \normal_{ij} \cdot \normal_{j} \frac{ A_{j}}{V_{i}} \\
        \jacobian_{FE,ji} &= c\normal_{j} \cdot \left(\frac{\normal_{ij}}{3\Delta x_{j}} + \gamma_{ij,n} \right).
        \end{align}
        The boundary Jacobian is (also with off-diagonal parts)
        \begin{equation}
        \jacobian_{FE,ji}^\mathrm{bc} = -\left(1-\alpha_{j}\right) \kappa^\mathrm{out}_{j,t}.
        \end{equation}
    \end{subequations}

\subsection{Particle corner case} \label{sec:appendix:particle_corner}

    Special caution is necessary if a particle leaves a cell at a corner. This is indicated by two surfaces with the same distance. Even though this case seems unlikely, it has been observed during calculations. If a particle crosses a cell corner, the next cell it enters is ambiguous. If not handled correctly, the particle can enter a cell without actually being within the cell boundaries. Our mitigation strategy is to move the intersection into one of the connected surfaces, resolving all ambiguity. Due to floating-point round-off issues, this is triggered if the particle crosses a surface in a small circle around a corner, with a fraction of the distance from the surface to the cell center as radius. Our solution addresses round-off issues and is valid for all mesh sizes. 

\subsection{Negative temperatures} \label{sec:appendix:negative_temperatures}

    A linear source representation can lead to negative values. To prevent this we check all corners of a cell, and adjust, if necessary the slope. For all vertices \(v\) of cell \(i\), perform the test
    \begin{equation}
    \avg{\Theta}_{i,n} + \linearv{\Theta}_{i,n} \cdot \left(\pos_{v} - \ccenter \right) < \Theta_{\min}
    \end{equation}
    where \(\Theta_{\min}\) is the lowest allowed temperature. If the inequality is true, we will adjust the slope such that the value at the vertex is \(\Theta_{\min}\). For this, we use the equation
    \begin{equation}
    \avg{\Theta}_{i,n} + \beta_v \linearv{\Theta}_{i,n} \cdot  \left(\pos_{v} - \ccenter \right)  = \Theta_{\min}
    \end{equation}
    and solve for 
    \begin{equation}
    \beta_v = \frac{ \Theta_{\min} - \avg{\Theta}_{i,n}}{\linearv{\Theta}_{i,n} \vd  \left(\pos_{v} - \ccenter \right)}.
    \end{equation}
    The slope is corrected as
    \begin{equation}
    \linearv{\Theta}'_{i,n} = \beta_v\linearv{\Theta}_{i,n}
    \end{equation}
    This will reduce the slope so that all corners with the negative values are reset to the minimum temperature or above, while preserving the cell average.

\end{appendices}
\end{document}